\documentclass[ste,twoside]{stefano}

\usepackage{amsmath}
\usepackage{amssymb}
\usepackage{graphics}
\usepackage{rotating}
\usepackage{cite}

\def\makeheadbox{{%
\hbox to0pt{\vbox{\baselineskip=10dd\hrule\hbox
to\hsize{\vrule\kern3pt\vbox{\kern3pt
\hbox{  {\sf Journal Physics A} {\bf 33}, 2971--2995 
(2000)}
\hbox{  {\sf math-ph/0002051} \hspace*{11.3cm} 
$\boldsymbol{\Sigma \delta \Lambda}$ }
\kern3pt}\hfil\kern3pt\vrule}\hrule}%
\hss}}}

\def\u{\leavevmode\hbox{\normalsize1\kern-4.2pt\large1}}
\def\0{\mbox{\tiny $0$}}
\def\1{\mbox{\tiny $1$}}
\def\2{\mbox{\tiny $2$}}
\def\3{\mbox{\tiny $3$}}
\def\4{\mbox{\tiny $4$}}
\def\5{\mbox{\tiny $5$}}
\def\8{\mbox{\tiny $8$}}
\def\m{\mbox{\tiny $m$}}
\def\n{\mbox{\tiny $n$}}
\def\l{\mbox{\tiny $l$}}
\def\i{\mbox{\tiny $i$}}
\def\d{\mbox{\tiny $diag$}}
\def\dn{\mbox{\tiny $2n$}}

\def\nm{\mbox{\small $n \times n$}}
\def\dnm{\mbox{\small $2n \times 2n$}}
\def\od{\mbox{\footnotesize $2 \times 2$}}

\def\-{\mbox{\tiny $-$}}
\def\+{\mbox{\tiny $+$}}
\def\={\mbox{\tiny $=$}}
\def\ll{\mbox{\tiny ${\lambda}$}}
\def\lc{\mbox{\tiny ${\lambda}^*$}}
\def\la{\mbox{\tiny $\lambda_{\1}$}}
\def\lac{\mbox{\tiny ${\lambda}_{\1}^*$}}
\def\li{\mbox{\tiny $\lambda_{\l}$}}
\def\lic{\mbox{\tiny ${\lambda}_{\l}^*$}}
\def\ln{\mbox{\tiny $\lambda_{\n}$}}
\def\lnc{\mbox{\tiny ${\lambda}_{\n}^*$}}
\def\lae{\mbox{\tiny $(\lambda_{\1})$}}
\def\laec{\mbox{\tiny $({\lambda}_{\1}^*)$}}
\def\lme{\mbox{\tiny $(\lambda_{\m})$}}

\def\lne{\mbox{\tiny $(\lambda_{\n})$}}
\def\lnec{\mbox{\tiny $({\lambda}_{\n}^*)$}}
\def\lb{\mbox{\tiny $\lambda_{\2}$}}
\def\lbc{\mbox{\tiny ${\lambda}_{\2}^*$}}
\def\ldn{\mbox{\tiny $\lambda_{\dn}$}}
\def\lde{\mbox{\tiny $(\lambda_{\2})$}}
\def\ldnae{\mbox{\tiny $(\lambda_{\dn - \1})$}}
\def\ldne{\mbox{\tiny $(\lambda_{\dn})$}}

\def\L{\mbox{\tiny $L$}}
\def\R{\mbox{\tiny $R$}}

\def\sdag{\mbox{\tiny $\dag$}}

\def\spsi{\mbox{\tiny $\psi$}}
\def\sphi{\mbox{\tiny $\phi$}}
\begin{document}
%
%%%%%%%%%%%%%%%%%%%%%%%%%%%%%%%% PAPER %%%%%%%%%%%%%%%%%%%%%%%%%%%%%%%%%%%%%

\title{Right eigenvalue equation in quaternionic quantum mechanics}

\author{
Stefano De Leo\inst{1}
\and
Giuseppe Scolarici\inst{2}
}

\institute{
Department of Applied Mathematics, University of Campinas\\
PO Box 6065, SP 13083-970, Campinas, Brazil\\
{\em deleo@ime.unicamp.br}
\and
Department of Physics, University of Lecce\\
PO Box 193, I 73100, Lecce, Italy\\
{\em scolarici@le.infn.it}
} 

\date{September 10, 1999}

%%%%%%%%%%%%%%%%%%%%%%%%%%%%%%%%%%%%%%%%%%%%%%%%%%%%%%%%%%%%%%%%%%%%%%%%%%%%%%%
%                                ABSTRACT
%%%%%%%%%%%%%%%%%%%%%%%%%%%%%%%%%%%%%%%%%%%%%%%%%%%%%%%%%%%%%%%%%%%%%%%%%%%%%%%

\abstract{
We study the right eigenvalue equation for quaternionic and 
complex linear matrix operators defined in $n$-dimensional quaternionic 
vector spaces. For quaternionic linear operators the eigenvalue spectrum  
consists of $n$ complex values. For these operators we give a necessary and 
sufficient condition for the  diagonalization of their quaternionic matrix 
representations. Our discussion is also extended to 
complex linear operators, whose spectrum is  characterized by $2n$ complex 
eigenvalues. We show that a consistent analysis   
of the eigenvalue problem for complex linear operators requires the choice
of a complex geometry in defining inner products. Finally, we introduce 
some examples of the left  eigenvalue equations and highlight the main 
difficulties in their solution.}

%%%%%%%%%%%%%%%%%%%%%%%%%%%%%%%%%%%%%%%%%%%%%%%%%%%%%%%%%%%%%%%%%%%%%%%%%%%%% 
%%%%%%%%%%%%%%%%%%%%%%%%%%%%%%%%%%%%%%%%%%%%%%%%%%%%%%%%%%%%%%%%%%%%%%%%%%%%%


%%%%%%%%%%%%%%%%%%%%%%%%%%%%%%%%%%%%%%%%%%%%%%%%%%%%%%%%%%%%%%%%%%%%%%%%%%%%%%%

\maketitle

%%%%%%%%%%%%%%%%%%%%%%%%%%%%%%%%%%%%%%%%%%%%%%%%%%%%%%%%%%%%%%%%%%%%%%%%%%%%%%%
%                                SECTION I
%%%%%%%%%%%%%%%%%%%%%%%%%%%%%%%%%%%%%%%%%%%%%%%%%%%%%%%%%%%%%%%%%%%%%%%%%%%%%%%
%%%%%%%%%%%%%%%%%%%%%%%%%%%%%%%%%%%%%%%%%%%%%%%%%%%%%%%%%%%%%%%%%%%%%%%%%%%%%%%

\section{Introduction}
\label{s1}

In the last decade, after the fundamental works of 
Finkelstein et al.~\cite{FIN1,FIN2,FIN3}  
on foundations of quaternionic Quantum Mechanics [qQM] 
and gauge theories, we have 
witnessed a renewed interest in algebrization and geometrization of
physical theories by non commutative fields~\cite{DIX,GUR}. 
Among the numerous references on this subject, we recall the important paper of
Horwitz and Biedenharn~\cite{HOR}, where the authors showed that the
assumption of a complex projection of the scalar product, also called 
{\em complex geometry}~\cite{REM}, 
permits the definition of a suitable tensor product~\cite{DEL1} 
between single-particle quaternionic wave functions. We also mention 
quaternionic applications in special relativity~\cite{DEL2}, group
representations~\cite{ADL,SCO,SCO2,DEL3}, 
non relativistic~\cite{ADL1,DAV} and relativistic dynamics~\cite{Rel1,Rel2}, 
field theory~\cite{FT}, lagrangian
formalism~\cite{DELL},    
electroweak model~\cite{DEL4}, grand  
unification theories~\cite{DEL5}, preonic model~\cite{ADL2}.
A clear and detailed discussion of qQM together possible topics for future 
developments in field theory and particle physics is found in the recent 
book of Adler~\cite{ADL3}. 

In writing this paper, the main objective  has been 
to address lack of clarity among mathematical physicists on  
the proper choice of the quaternionic eigenvalue 
equation within a qQM with {\em complex} or {\em quaternionic geometry}.
In the past, interesting  papers have addressed the mathematical discussion of 
the quaternionic eigenvalue equation, and related topics.  
For example, we find in literature works on quaternionic 
eigenvalues and
characteristic equation~\cite{LEE,BRE}, diagonalization  
of matrices~\cite{COH},   
Jordan form and q-determinant~\cite{DET,DNIR}.    
More recently, some of these problems  have been also discussed for 
the octonionic field~\cite{DM1,DM2}. 

Our approach  aims to give a practical method in solving the quaternionic 
right  eigenvalue equation in view of rising interest in 
quaternionic~\cite{GUR,ADL3,DR}
and octonionic~\cite{DIX,OCT1,OCT2,DK1,DK2,OCT3} applications in physics.
Given quaternionic and complex linear operators on $n$-dimensional 
quaternionic vector spaces, we explicitly formulate 
practical rules to obtain  the eigenvalues and the corresponding eigenvectors 
for their $n$-dimensional 
quaternionic matrix representations. In discussing the right complex 
eigenvalue problem in qQM, we find two 
obstacles. The first one is related to the difficulty in obtaining a 
suitable definition of determinant for 
quaternionic matrices, the second one is represented by the loss, 
for non-commutative fields, of the fundamental theorem of the algebra.  
The lack of these tools, essentials in solving the 
eigenvalue problem in the complex world, make the problem over the 
quaternionic field a complicated puzzle. We overcome the difficulties in 
approaching the eigenvalue problem in a quaternionic world by 
discussing the eigenvalue equation for $2n$-dimensional complex matrices
obtained by translation from $n$-dimensional  quaternionic matrix operators.
We shall show that 
quaternionic  linear operators, defined 
on quaternionic Hilbert space with quaternionic geometry, are diagonalizable 
if and only if the corresponding complex operators are diagonalizable.  
The spectral theorems, extended to quaternionic Hilbert spaces~\cite{FIN1}, 
are recovered in a more general context. 
We also study the linear independence on $\mathbb{H}$  of quaternionic 
eigenvectors by studying their (complex) eigenvalues and discuss 
the spectrum choice for quaternionic quantum systems.
Finally, we construct the hermitian operator associated with 
any anti-hermitian matrix operator and show that a coherent  
discussion of the eigenvalue problem for complex linear operators
operators requires a complex geometry. A brief discussion concerning 
the possibility to have left eigenvalue equations is also
proposed and some examples presented. We point out, see also~\cite{DM1,OCT3}, 
that left eigenvalues of hermitian quaternionic matrices need not be
real. 

This paper is organized as follows: In section~\ref{s2}, we introduce  
basic notations and mathematical tools. In particular, we discuss 
similarity transformations, symplectic decompositions  and the left/right 
action of quaternionic imaginary units. In section~\ref{s2b}, we give 
the basic framework of qQM and translation rules between complex and
quaternionic matrices. In section~\ref{s3}, we approach the
 right eigenvalue problem by discussing the eigenvalue spectrum for 
$2n$-dimensional complex matrices obtained by translating 
$n$-dimensional quaternionic matrix representations. We give a 
practical method to diagonalize $n$-dimensional quaternionic linear 
matrix operators and overcome previous problems in the spectrum choice for
quaternionic quantum systems. 
We also discuss the right eigenvalue problem for complex
linear operators within a qQM with
 complex geometry. In section~\ref{s4}, we introduce 
the left eigenvalue equation and analyze the eigenvalue spectrum for hermitian
operators. We  explicitly solve some examples of left/right 
eigenvalue equations for two-dimensional quaternionic and complex linear  
operators.  Our conclusions and out-looks are drawn in the final section.

%%%%%%%%%%%%%%%%%%%%%%%%%%%%%%%%%%%%%%%%%%%%%%%%%%%%%%%%%%%%%%%%%%%%%%%%%%%%%%
%%%%%%%%%%%%%%%%%%%%%%%%%%%%%%%%%%%%%%%%%%%%%%%%%%%%%%%%%%%%%%%%%%%%%%%%%%%%%%%
%                           SECTION 2
%%%%%%%%%%%%%%%%%%%%%%%%%%%%%%%%%%%%%%%%%%%%%%%%%%%%%%%%%%%%%%%%%%%%%%%%%%%%%%%
%%%%%%%%%%%%%%%%%%%%%%%%%%%%%%%%%%%%%%%%%%%%%%%%%%%%%%%%%%%%%%%%%%%%%%%%%%%%%%%
\section{Basic notations and mathematical tools}
\label{s2}

A quaternion, 
$q \in \mathbb{H}$, is expressed by four real quantities~\cite{HAM,HAM2}
\begin{equation}
q = a + ib + jc + kd~,~~~~~a,b,c,d \in \mathbb{R}
\end{equation}
and three imaginary units
\[
i^{\2} = j^{\2} = k^{\2} = ij k=-1~.
\]
The quaternion skew-field $\mathbb{H}$ is an associative but non-commutative 
algebra of rank 4  over $\mathbb{R}$, endowed with an involutory 
antiautomorphism 
\begin{equation}
q \rightarrow \overline{q} = a - ib - jc - kd~.
\end{equation}
This conjugation implies a reversed order product, namely
\[
 \overline{pq} = \overline{q} \, \overline{p}~,~~~~~p~,~q \in \mathbb{H}.
\]
Every nonzero quaternion is invertible, and the unique inverse is given by 
$1/q = \overline{q} / | q |^{\2}$, where the quaternionic norm $|q|$ is defined
by 
\[
| q |^{\2} = q \overline{q} = a^{\2} + b^{\2} + c^{\2} + d^{\2}~.  
\]

\noindent $\bullet$ {\bf Similarity Transformation}

\noindent 
Two quaternions $q$ and $p$ belong to the same eigenclass when the following 
relation 
\[
q = s^{\- \1} \, p\, s~,~~~~~
s \in \mathbb{H}~,
\]
is satisfied. Quaternions of the same eigenclass 
have the same real part and the same norm, 
\[ 
\mbox{Re}(q)=
\mbox{Re} \left( s^{\- \1} \, p \, s \right)=
\mbox{Re}(p)~,~~~~~
|q| = | s^{\- \1} \, p \, s | = |p|~,
\]
consequently they have the same absolute value of the imaginary part. 
The previous equations can be rewritten in terms of unitary quaternions
as follows
\begin{equation}
\label{stra}
q = s^{\- \1} \, p \, s 
  = \frac{\overline{s}}{|s|} \, p \, \frac{s}{|s|}
  = \overline{u} \, p \, u~,~~~~~u \in \mathbb{H}~,~~~
\overline{u} u = 1~.
\end{equation}
In eq.~(\ref{stra}), the unitary quaternion,
\[
u = \cos \mbox{$\frac{\theta}{2}$} + \vec{h} \cdot \vec{u} \, 
\sin \mbox{$\frac{\theta}{2}$}~,~~
~~~\vec{h} \equiv (i,j,k)~,~\vec{u} \in \mathbb{R}^{\3}~,~
|\vec{u} \, | = 1~,
\]
can be expressed in terms of the imaginary parts of 
$q$ and $p$. In fact, given two quaternions belong to the same 
eigenclass,
\[
q = q_{\0} + \vec{h} \cdot \vec{q}~,~~~
p = p_{\0} + \vec{h} \cdot \vec{p}~,~~~~~
q_{\0}=p_{\0}~,~~~|\vec{q} \, | = |\vec{p} \, |~,
\]
we find~\cite{PHD}, for $\vec{q} \neq \pm \, \vec{p}$,
\begin{equation}
\label{GIS}
\cos \theta = \frac{\vec{q} \cdot \vec{p}}{|\vec{q}| \, |\vec{p}|}~~~~~
\mbox{and}~~~~~
\vec{u} = \frac{\vec{q} \times \vec{p}}{|\vec{q} \, | \, |\vec{p} \, | \, 
\sin \theta}~.
\end{equation}
The remaining cases $\vec{q} = \vec{p}$ and $\vec{q} = - \, \vec{p}$  
represent respectively the trivial similarity transformation, 
i.e. $uq \bar{u}=q$, and the similarity transformation
between a quaternion $q$ and its conjugate $\bar{q}$, i.e. 
$u q \bar{u} = \bar{q}$. In the first case the unitary quaternion is given
by the identity quaternion. In the last case the similarity transformation
is satisfied $\forall u=\vec{h} \cdot \vec{u}$    
with $\vec{u} \cdot \vec{q}=0$ and $| \vec{u} \,|=1$.\\

\noindent $\bullet$ {\bf Symplectic Decomposition}

\noindent Complex numbers can be constructed from real numbers by 
\[ 
z = \alpha + i \beta~,~~~~~ \alpha , \beta \in \mathbb{R}~.
\]
In a similar way, we can construct quaternions from complex numbers by 
\[ 
q =  z + jw~,~~~~~z,w \in \mathbb{C}~,
\]
{\em symplectic decomposition} of quaternions.

\noindent $\bullet$ {\bf Left/Right Action}

\noindent 
Due to the non-commutative nature of quaternions we must distinguish between 
\[ q \vec{h}~~~~~\mbox{and}~~~~~\vec{h}q~.\]
Thus,  it is appropriate to consider left and
right-actions for our imaginary units $i$, $j$ and $k$. 
Let us define the operators
\begin{equation}
\vec{L} = \left(  L_i , L_j , L_k  \right),
\end{equation} 
and
\begin{equation}
\vec{R} = \left(  R_i , R_j , R_k  \right),
\end{equation} 
which act on quaternionic states in the following way
\begin{equation}
\vec{L} : ~\mathbb{H} \rightarrow \mathbb{H}~,~~~~~     
\vec{L} q = \vec{h} q \in \mathbb{H}~,
\end{equation}
and
\begin{equation}
\vec{R} : ~\mathbb{H} \rightarrow \mathbb{H}~,~~~~~     
\vec{R} q = q \, \vec{h} \in \mathbb{H}~.
\end{equation}
The algebra of left/right generators can be concisely expressed by 
\[ 
L_i^{\2} = L_j^{\2} =L_k^{\2} =L_i L_j L_k =
R_i^{\2} = R_j^{\2} =R_k^{\2} =R_k R_j R_i =
- \u~,
 \]
and by the commutation relations
\[ 
\left[ \, L_{i,j,k} \, , \, R_{i,j,k} \, \right] = 0~.
\]
From these operators we can construct the following vector space
\[
\mathbb{H}_{\L} \otimes \mathbb{H}_{\R},
\]
whose generic element will be characterized by left and right actions of
quaternionic imaginary units $i$, $j$, $k$. 
In this paper we will work with two sub-spaces of
$\mathbb{H}_{\L} \otimes \mathbb{H}_{\R}$, namely
\[ \mathbb{H}^{\L}~~~~~\mbox{and}~~~~~
  \mathbb{H}^{\L} \otimes  \mathbb{C}^{\R}~,
\]
whose elements are represented respectively by
left actions of $i$, $j$, $k$ 
\begin{equation}
\label{n1} 
a + \vec{b} \cdot \vec{L}~~ \in \mathbb{H}^{\L}~,~~~~~
a, \vec{b} \in \mathbb{R}~,
\end{equation}
and by left actions of $i$, $j$, $k$  and right action of the only
imaginary unit $i$
\begin{equation}
\label{n2}
a + \vec{b} \cdot \vec{L} + c R_i + \vec{d} \cdot \vec{L} R_i~~ 
\in \mathbb{H}^{\L} \otimes \mathbb{C}^{\R}~,~~~~~
a,\vec{b},c,\vec{d} \in \mathbb{R}~.
\end{equation}

%%%%%%%%%%%%%%%%%%%%%%%%%%%%%%%%%%%%%%%%%%%%%%%%%%%%%%%%%%%%%%%%%%%%%%%%%%%%%%
%%%%%%%%%%%%%%%%%%%%%%%%%%%%%%%%%%%%%%%%%%%%%%%%%%%%%%%%%%%%%%%%%%%%%%%%%%%%%%%
%                           SECTION 2bis
%%%%%%%%%%%%%%%%%%%%%%%%%%%%%%%%%%%%%%%%%%%%%%%%%%%%%%%%%%%%%%%%%%%%%%%%%%%%%%%
%%%%%%%%%%%%%%%%%%%%%%%%%%%%%%%%%%%%%%%%%%%%%%%%%%%%%%%%%%%%%%%%%%%%%%%%%%%%%%%
\section{States and Operators in qQM}
\label{s2b}

The states of qQM will be described by vectors, $| \psi \rangle$, of a 
quaternionic Hilbert space, $V_{\mathbb{H}}$. 
First of all, due to the non-commutative nature of quaternionic multiplication,
 we must specify whether the quaternionic Hilbert space is to be formed by 
right or left multiplication of quaternionic vectors by scalars. 
The two different
 conventions give isomorphic versions of the theory~\cite{IVT}.
We adopt the convention of right multiplication by scalars.

In quaternionic Hilbert spaces, we can define quaternionic and complex
linear operators, which will be respectively denoted by 
${\cal O}_{\mathbb{H}}$ and ${\cal O}_{\mathbb{C}}$. They will act on 
quaternionic vectors, $| \psi \rangle$, in the following way 
\[
{\cal O}_{\mathbb{H}}  ( | \psi \rangle q ) = 
\left( {\cal O}_{\mathbb{H}} | \psi \rangle \right) q~,~~~~~
q \in  \mathbb{H}~.
\]
and 
\[
{\cal O}_{\mathbb{C}}  ( | \psi \rangle \lambda ) = 
\left( {\cal O}_{\mathbb{C}} | \psi \rangle \right) \lambda~,~~~~~
\lambda \in  \mathbb{C}~.
\] 
Such operators are $\mathbb{R}$-linear from the left.

As a concrete illustration, let us consider the case of a finite 
$n$-dimensional quaternionic Hilbert space. The ket state $| \psi \rangle$
will be represented by a quaternionic $n$-dimensional column vector
\begin{equation}
|\psi \rangle  = 
\left( \begin{array}{c} \psi_{\1} \\ \vdots \\ 
 \psi_{\n}  \end{array} \right) = 
\left( \begin{array}{c} x_{\1} + j y_{\1} \\ \vdots \\ 
 x_{\n} + j y_{\n}  \end{array} \right)~,~~~~~
x_{\1},y_{\1}, ... , x_{\n}, y_{\n} \in \mathbb{C}~.
\end{equation}
Quaternionic linear operators, $\mathcal{O}_{\mathbb{H}}$, will be represented
by $\nm$ matrices with entries in $\mathbb{H}^{\L}$, whereas complex linear
operators, $\mathcal{O}_{\mathbb{C}}$, by $\nm$ matrices with entries in 
$\mathbb{H}^{\L} \otimes \mathbb{C}^{\R}$.

By using the {\em symplectic} 
complex representation, the $n$-dimensional quaternionic vector
\[ 
| \psi \rangle = |x\rangle + j \, |y\rangle =
\left( \begin{array}{c} x_{\1} \\ \vdots \\ 
 x_{\n} \end{array} \right) + j \, 
\left( \begin{array}{c} y_{\1} \\ \vdots \\ 
 y_{\n}  \end{array} \right)~,
\]
can be translated in the  $2n$-dimensional complex column vector
\begin{equation}
|\psi \rangle  \leftrightarrow  
\left( \begin{array}{c} x_{\1} \\ y_{\1} \\ \vdots \\ 
 x_{\n} \\ y_{\n}  \end{array} \right)~.
\end{equation}

The matrix  representation of 
$L_i$, $L_j$ and $L_k$ consistent with the above identification is
\begin{equation}
\label{ident}
   L_i \leftrightarrow \left( \begin{array}{cc} i & 0\\ 0 & $-$ i
                              \end{array} \right)=i\sigma_{\3}~,~~~~~
   L_j \leftrightarrow \left( \begin{array}{cc} 0 & $-$ 1\\ 1 & 0
                              \end{array} \right)= - i\sigma_{\2}~,~~~~~
   L_k \leftrightarrow \left( \begin{array}{cc} 0 &  $-$ i\\ $-$ i & 0
                              \end{array} \right)= - i\sigma_{\1}~.
\end{equation}
These translation rules allow to represent quaternionic $n$-dimensional 
linear operators by $\dnm$ complex matrices. 

The right quaternionic imaginary unit
\begin{equation}
\label{tri}
 R_i ~\leftrightarrow ~ \left( \begin{array}{cc} i & 0\\ 0 & i
                                       \end{array} \right)~,
\end{equation}
adds four additional degrees of freedom, 
\[ R_i~,~L_i R_i~,~L_j R_i~,~L_k R_i~,\] 
and so, by observing that the Pauli matrices together with the identity matrix
form a basis over $\mathbb{C}$, 
we have a set of rules which allows to translate one-dimensional complex
linear operators by $\od$ complex matrices~\cite{DEL6}. Consequently, we
can construct $\dnm$ complex matrix representations for 
$n$-dimensional complex linear operators.

Let us note that the identification in  eq.~(\ref{tri}) is consistent only
for complex inner products (complex geometry)~\cite{DEL6}. We observe that 
the right complex imaginary 
unit, $R_{\i}$, has not a well defined 
hermiticity within a qQM with quaternionic inner product 
(quaternionic geometry),
\begin{equation}
\label{qgeo}
\langle \varphi | \psi \rangle = 
| \varphi \rangle^{\sdag} | \psi \rangle = 
\sum_{l=1}^{n} 
\bar{\varphi}_{\l} \psi_{\l}~.
\end{equation}
In fact, anti-hermitian operators must satisfy
\[
\langle \varphi| {\cal A} \psi\rangle  =
- \langle {\cal A} \varphi | \psi \rangle~.
\]
For the right imaginary unit $R_{\i}$, we have
\[
| R_{\i} \psi \rangle \equiv R_{\i} | \psi \rangle = 
| \psi \rangle i~,~~~
\langle  R_{\i} \varphi | \equiv  
\left( R_{\i} | \varphi \rangle \right)^{\sdag} = 
- i \langle  \varphi | ~,
\]
and consequently
\[ 
\langle \varphi | \psi \rangle i = 
\langle \varphi | R_{\i} \psi \rangle \neq  
- \langle  R_{\i} \varphi |\psi \rangle  = i 
\langle \varphi | \psi \rangle~.
\] 
Nevertheless,  
by adopting a complex geometry, i.e. a complex projection of the quaternionic 
inner product, 
\begin{equation}
\label{cgeo}
\langle \varphi | \psi \rangle_{\mathbb{C}} = 
\frac{ \langle \varphi | \psi \rangle - i  
\langle \varphi | \psi \rangle i}{2}~, 
\end{equation}
we recover the anti-hermiticity of the operator $R_{\i}$,
\[
\langle \varphi| R_{\i} \psi\rangle_{\mathbb{C}}  =
- \langle R_{\i} \varphi | \psi \rangle_{\mathbb{C}}~. 
\]

%%%%%%%%%%%%%%%%%%%%%%%%%%%%%%%%%%%%%%%%%%%%%%%%%%%%%%%%%%%%%%%%%%%%%%%%%%%%%%%
%%%%%%%%%%%%%%%%%%%%%%%%%%%%%%%%%%%%%%%%%%%%%%%%%%%%%%%%%%%%%%%%%%%%%%%%%%%%%%%
%                              SECTION 3
%%%%%%%%%%%%%%%%%%%%%%%%%%%%%%%%%%%%%%%%%%%%%%%%%%%%%%%%%%%%%%%%%%%%%%%%%%%%%%%
%%%%%%%%%%%%%%%%%%%%%%%%%%%%%%%%%%%%%%%%%%%%%%%%%%%%%%%%%%%%%%%%%%%%%%%%%%%%%%%

\section{The right complex eigenvalue problem in qQM }
\label{s3}

The right eigenvalue equation for a generic quaternionic linear operator, 
$O_{\mathbb{H}}$, is written as
\begin{equation}
\label{ev}
O_{\mathbb{H}} | \Psi \rangle\,= | \Psi \rangle q~,
\end{equation}
where $
| \Psi \rangle\, \in V_{\mathbb{H}}$ and $q \in \mathbb{H}$. 
By adopting quaternionic scalar products in our quaternionic Hilbert spaces,
$V_{\mathbb{H}}$, we find  states in one to one correspondence with unit rays 
of the form
\begin{equation}
\label{de}
| \mathbf{r} \rangle \, = \{|\Psi\rangle u \}
\end{equation}
where $|\Psi\rangle$ is a normalized vector and $u$ a quaternionic 
phase of magnitude unity. The state vector, $|\Psi\rangle u$, corresponding
 to the same physical state $|\Psi\rangle$, is an $O_{\mathbb{H}}$-eigenvector
with eigenvalue $\overline{u} q u$
\[  
O_{\mathbb{H}} | \Psi \rangle u  =
| \Psi \rangle u \, (\overline{u} q u)~.
\]
For real values of $q$, we find only one eigenvalue, otherwise 
quaternionic linear operators will be  
characterized by an infinite eigenvalue spectrum
\[
\{q~,~\overline{u}_{\1} q u_{\1}~,~...~,~
\overline{u}_{\l} q u_{\l}~,~...\}
\]
with $u_{\l}$ unitary quaternions.
The related set of eigenvectors 
\[
\biggl\{| \Psi \rangle~,~| \Psi \rangle u_{\1}~,~...~,~| \Psi \rangle u_{\l}~,~
...\biggr\}
\]
represents a ray. We can characterize our spectrum by 
choosing a representative ray
\[
|\psi\rangle = |\Psi\rangle u_{\ll}~,
\]
so that the corresponding  eigenvalue 
$\lambda = \bar{u}_{\ll} q  u_{\ll}$ is complex. 
For this state the right eigenvalue equation becomes
\begin{equation}
O_{\mathbb{H}} | \psi \rangle \, =
| \psi \rangle\lambda~,
\end{equation}
with $|\psi\rangle\, \in V_{\mathbb{H}}$ and $\lambda \in \mathbb{C}$.

We now give a systematic method to determine the complex eigenvalues
of quaternionic matrix representations for $O_{\mathbb{H}}$ operators.

%%%%%%%%%%%%%%%%%%%%%%%%%%%%%%%%%%%%%%%%%%%%%%%%%%%%%%%%%%%%%%%%%%%%%
%                      SUBSECTION 3.1
%%%%%%%%%%%%%%%%%%%%%%%%%%%%%%%%%%%%%%%%%%%%%%%%%%%%%%%%%%%%%%%%%%%%%

\subsection{Quaternionic linear operators and quaternionic geometry}

In $n$-dimensional quaternionic vector spaces, $\mathbb{H}^{\, \n}$, 
 quaternionic linear operators, $O_{\mathbb{H}}$, are represented by
 $n\times n$ quaternionic matrices, $\mathcal{M}_{\n}(\mathbb{H}^{\L})$,
 with elements
in $\mathbb{H}^{\L}$.
Such quaternionic matrices admit $2n$-dimensional 
complex counterparts by the translation rules given in eq.~(\ref{ident}).
Such  complex matrices characterize
a subset of the $2n$-dimensional complex matrices
\[ 
\widetilde{M}_{\dn}(\mathbb{C}) \subset M_{\dn}(\mathbb{C})~.
\]
The eigenvalue equation for $O_{\mathbb{H}}$ reads
\begin{equation}
\label{mev}
 \mathcal{M}_{\mathbb{H}} | \psi \rangle \, = | \psi \rangle \lambda~,
\end{equation}
where $
\mathcal{M}_{\mathbb{H}}\in \mathcal{M}_{n}(\mathbb{H^{\L}})$, $
| \psi \rangle \, \in \mathbb{H}^{\, \n}$ and $\lambda \in \mathbb{C}$.
%%%%%%%%%%%%%%%%%%%%%%%%%%%%%%%%%%%%%%%%%%%%%%%%%%%%%%%%%%%%%%%%%%%%%%%%%%%%
%%%%%%%%%%%%%%%%%%%%%%%%%%%%%%%%%%%%%%%%%%%%%%%%%%%%%%%%%%%%%%%%%%%%%%%%%%%%

\subsection*{$\bullet$ The one-dimensional eigenvalue problem}

In order to introduce the reader to our general method of quaternionic matrix
diagonalization, let us discuss one-dimensional right complex eigenvalue
equations.
In this case eq.~(\ref{mev}) becomes
\begin{equation}
\label{a}
Q_{\mathbb{H}} | \psi \rangle\,=
 | \psi \rangle \lambda~,
\end{equation}
where $
Q_{\mathbb{H}} = a + \vec{b} \cdot \vec{L}~ \in \mathbb{H}^{\L}$, 
$|\psi\rangle\,= |x\rangle + j \, |y\rangle\, \in \mathbb{H}$ and 
$\lambda \in \mathbb{C}$. 
By using the translation rules, given in section~\ref{s2b}, we can generate
the quaternionic algebra from the commutative complex algebra 
(Cayley-Dickson process). The 
complex counterpart of eq.(\ref{a}) reads 
\begin{equation}
\label{d}
\begin{pmatrix}
 z & $-$ w^{\ast} \\ 
w & z^{\ast} 
\end{pmatrix} 
\begin{pmatrix}
 x\\ y 
\end{pmatrix} =
\lambda
\begin{pmatrix} 
x\\ y 
\end{pmatrix}~,
\end{equation}
\[
z = a + ib_{\1}~,~~~ w = b_{\2} - ib_{\3} \in \mathbb{C}~.
\]
Eq.~(\ref{d}) is the eigenvalue equation for a complex matrix whose 
characteristic equation has real coefficients. For this reason,
the translated complex operator admits $\lambda$ and $\lambda^{\ast}$ as 
eigenvalues. 

Given the eigenvector corresponding to the 
eigenvalue $\lambda$, we can immediately obtain the eigenvector associated
to the eigenvalue $\lambda^{\ast}$  by taking the complex
conjugate of eq.(\ref{d}) and then applying a similarity transformation
by the matrix
\[
S=
\begin{pmatrix}
 0 & $-$1 \\ 1 & 0
\end{pmatrix}.
\]
In this way, we find
\begin{gather}
\label{f}
\begin{pmatrix}
 z & $-$w^{\ast}
\\ w & z^{\ast}
\end{pmatrix}
\begin{pmatrix}
 $-$y^{\ast}\\
 x^{\ast}
\end{pmatrix}=
\lambda^{\ast}
\begin{pmatrix}
 $-$y^{\ast} \\
 x^{\ast}
\end{pmatrix}.
\end{gather}
So, for $\lambda \neq \lambda^{\ast} \in \mathbb{C}$,  
we obtain the eigenvalue spectrum $\{\lambda \, , \lambda^{\ast}\}$ 
with eigenvectors
\begin{equation}
\label{eg}
\begin{pmatrix} 
x\\y
\end{pmatrix}~~~,~~~
\begin{pmatrix}
 $-$y^{\ast}\\x^{\ast}
\end{pmatrix}~.
\end{equation}

What happens when $\lambda \in \mathbb{R}$ ? In this case 
the eigenvalue spectrum will be  determined by two equal eigenvalues
 $\lambda$. To show that,  we remark that the eigenvectors~(\ref{eg}) 
associated to the same eigenvalues $\lambda$, are linearly independent
on $\mathbb{C}$. In fact,
\[
\begin{Vmatrix}
 x & $-$ y^{\ast} \\ y & x^{\ast} 
\end{Vmatrix} =
|x|^{\2} + |y|^{\2} = 0~~~~~
\mbox{if and only if}~
x = y = 0~.
\]
So in the quaternionic world, by translation, we find two complex 
eigenvalues respectively $\lambda$ and $\lambda^{\ast}$, associated
 to the following quaternionic eigenvectors
\[
 |\psi\rangle ~~~\mbox{and}~~~ |\psi\rangle j~~~\in ~| \mathbf{r} \rangle ~.
\]
The infinite quaternionic eigenvalue spectrum can be
 characterized by the complex eigenvalue $\lambda$ and the ray 
representative will be $|\psi\rangle $. 
In the next section, by using the same method, we will discuss eigenvalue
equations in $n$-dimensional quaternionic vector spaces.

%%%%%%%%%%%%%%%%%%%%%%%%%%%%%%%%%%%%%%%%%%%%%%%%%%%%%%%%%%%%%%%%%%%%%%%%%%%%
%%%%%%%%%%%%%%%%%%%%%%%%%%%%%%%%%%%%%%%%%%%%%%%%%%%%%%%%%%%%%%%%%%%%%%%%%%%%

\subsection*{$\bullet$ The $n$-dimensional eigenvalue problem}

Let us formulate two theorems which generalize
the previous results for quaternionic
 $n$-dimensional eigenvalue problems. The first theorem [T1] analyzes the
eigenvalue spectrum of the $2n$-dimensional complex matrix
 $\widetilde{M}$, the complex counterpart of the $n$-dimensional quaternionic 
matrix $\mathcal{M}_{\mathbb{H}}$. In this theorem we 
give the matrix $S$ which allows to construct the complex eigenvector  
$| \phi_{\lc} \rangle$  from the eigenvector  $| \phi_{\ll} \rangle$. 
The explicit construction of $| \phi_{\lc} \rangle$ enables us to show the 
linear independence on 
$\mathbb{C}$ of $|\phi_{\ll} \rangle$ and  $| \phi_{\lc} \rangle$ 
when $\lambda \in \mathbb{R}$ and  represents  the main tool in 
constructing similarity transformations for diagonalizable quaternionic 
matrices. The second theorem [T2] discusses linear independence on 
$\mathbb{H}$ for $\mathcal{M}_{\mathbb{H}}$-eigenvectors.

\noindent {\tt T1 - THEOREM}

 {\em Let $\widetilde{M}$ be the complex counterpart of a
 generic  $n\times n$ quaternionic matrix  $\mathcal{M}_{\mathbb{H}}$. Its
 eigenvalues appear in conjugate pairs.}

Let
\begin{equation}
\label{A}
\widetilde{M} \, |\phi_{\ll}\rangle  \, = \lambda \, |\phi_{\ll}\rangle   
\end{equation}
be the eigenvalue equation for $\widetilde{M}$, where
\[
\widetilde{M} \in M_{\dn} (\mathbb{C})~,~~~
|\phi_{\ll}\rangle  \, =
\begin{pmatrix}
 x_{\1}\\ y_{\1}  \\ \vdots \\ x_{\n}\\y_{\n}
 \end{pmatrix} ~
 \in \mathbb{C^{\, \dn}}~,~~~
 \lambda \in \mathbb{C}~.
\]
By taking the complex conjugate of eq.~(\ref{A}),
\[
\widetilde{M}^{\ast} | \phi_{\ll} \rangle ^{\ast}\,=\lambda^{\ast}
| \phi_{\ll} \rangle ^{\ast} ~,
\]
and applying a similarity transformation by the matrix
\[
S={\u}_{\n}\otimes 
\begin{pmatrix}
 0 & $-$1 \\ 1 & 0
 \end{pmatrix}~,
\]
we obtain
\begin{equation}
\label{st}
S \widetilde{M}^{\ast}S^{\-1}S | \phi_{\ll} \rangle ^{\ast}=\lambda^{\ast}
S | \phi_{\ll} \rangle ^{\ast}.
\end{equation}
From the block structure of the complex matrix $\widetilde{M}$
 it is easily checked that
\[
S\widetilde{M}^{\ast}S^{\-1}=
\widetilde{M},
\]
and consequently eq.~(\ref{st}) reads
\begin{equation}
\label{EQ}
\widetilde{M}\, | \phi_{\lc} \rangle \,=\lambda^{\ast}
| \phi_{\lc} \rangle ~,
\end{equation}
where
\[
| \phi_{\lc} \rangle  \, =
S | \phi_{\ll} \rangle ^{\ast} \, =
\begin{pmatrix} 
$-$y^{\ast}_{\1}\\ x^{\ast}_{\1} \\ \vdots \\
 $-$y^{\ast}_{\n}\\x^{\ast}_{\n}
 \end{pmatrix}.
\]
Let us show that the eigenvalues appear in conjugate pairs (this implies a 
double multiplicity for real eigenvalues). To do it, we need to prove
 that $|\phi_{\ll}\rangle $ and $|\phi_{\lc}\rangle $
 are  linearly independent on $\mathbb{C}$.
In order to demonstrate the linear independence of such eigenvectors,
 trivial for
 $\lambda \neq \lambda^{\ast}$, we observe that linear dependence, possible
 in the case $\lambda = \lambda^{\ast}$, should require
\[
\begin{Vmatrix}
 x_{i} & $-$ y_{i}^{\ast} \\ y_{i} & x_{i}^{\ast}
 \end{Vmatrix} =
 |x_{i}|^{2}+|y_{i}|^{2} = 0 \quad i=1,...,n~,
\]
verified only for null eigenvectors.
The linear independence of $|\phi_{\ll}\rangle $ and $|\phi_{\lc}\rangle $
ensures an even multiplicity for real eigenvalues $\blacksquare$.

We shall use the results of the first theorem to obtain informations
 about the $\mathcal{M}_{\mathbb{H}}$ right complex eigenvalue spectrum.
Due to non-commutativity nature of the quaternionic field we cannot 
give a suitable definition of determinant for quaternionic matrices 
and consequently we cannot write a characteristic polynomial $P(\lambda)$ 
for $\mathcal{M}_{\mathbb{H}}$. Another difficulty it is 
represented by the right position of the complex eigenvalue $\lambda$. 
 
\noindent {\tt T2 - THEOREM}

 {\em $\mathcal{M}_{\mathbb{H}}$ admits $n$ linearly independent 
eigenvectors on $\mathbb{H}$ if and only if its complex counterpart 
$\widetilde{M}$ admits $2n$ linearly independent eigenvectors on 
$\mathbb{C}$~.}

 Let
\begin{equation}
\label{F}
\biggl\{ | \phi_{\la} \rangle ~,~ | \phi_{\lac} \rangle ~,~...~,
~ | \phi_{\ln} \rangle ~,~| \phi_{\lnc} \rangle  \biggr\}
\end{equation}
be a set of $2n$ $\widetilde{M}$-eigenvectors, linearly independent 
on $\mathbb{C}$, and
$\alpha_{\l}$, $\beta_{\l}$ ($l = 1,...,n$) 
be generic complex coefficients. By definition
\begin{equation}
\label{E1}
\sum_{l=1}^{n}
 \biggl ( \alpha_{\l} | \phi_{\li} \rangle  + \, 
\beta_{\l} | \phi_{\lic} \rangle  \biggr ) = 0~~~
\Leftrightarrow~~~\alpha_{\l} = \beta_{\l} =0~.
\end{equation}
By translating the complex eigenvector set (\ref{F}) in quaternionic 
formalism we find
\begin{equation}
\label{G}
\biggl\{ |\psi_{\la}\rangle ~,~|\psi_{\lac}\rangle ~,~...~,
~|\psi_{\ln}\rangle ~,~|\psi_{\lnc}\rangle  \biggr\}.
\end{equation}
 By eliminating the eigenvectors, 
$|\psi_{\lic}\rangle  \, = |\psi_{\li}\rangle j$,
 corresponding for complex  eigenvalues to ones  
with negative  imaginary part, linearly dependent
 with $|\psi_{\li}\rangle $ on $\mathbb{H}$, we obtain
\[
\biggl\{ | \psi_{\la} \rangle  ~,~...~,~|\psi_{\ln}\rangle  \biggr\}.
\]
This set is formed by $n$ linearly independent vectors on $\mathbb{H}$.
In fact, by taking an arbitrary quaternionic linear combination of 
such vectors, we have 
\begin{equation}
\label{E2}
\sum_{l=1}^{n}
 \biggl[ |\psi_{\li}\rangle (\alpha_{\l} + j\beta_{\l}) \biggr] = 
\sum_{l=1}^{n} \biggl( |\psi_{\li}\rangle \alpha_{\l} +
 |\psi_{\lic}\rangle \beta_{\l} \biggr) = 0~~~
\Leftrightarrow~~~\alpha_{\l} = \beta_{\l} =0~.
\end{equation}
Note that eq.~(\ref{E2}) 
represents the quaternionic counterpart of eq.~(\ref{E1}) $\blacksquare$.

The $\mathcal{M}_{\mathbb{H}}$ complex eigenvalue spectrum is thus  
obtained by taking from the $2n$ dimensional $\widetilde{M}$-eigenvalues
 spectrum 
\[
\{ \lambda_{\1}~,~\lambda_{\1}^{\ast}~,~...~,~\lambda_{\n}~,
~\lambda_{\n}^{\ast} \}, 
\]
the reduced $n$-dimensional spectrum
\[
\{ \lambda_{\1}~,~...~,
~\lambda_{\n} \}~. 
\]
We stress here the fact that, the choice of positive, rather than negative,
 imaginary part is a simple convention. In fact, from the 
quaternionic eigenvector set (\ref{G}), we can extract different sets 
of quaternionic linearly independent eigenvectors
\[
\biggl\{ \left[ |\psi_{\la}\rangle ~ \mbox{or} ~|\psi_{\lac}\rangle  \right]~,~...~,~
\left[ |\psi_{\ln}\rangle ~ \mbox{or} ~|\psi_{\lnc}\rangle  \right] \biggr\}~,
\]
and consequently we have a free choice in characterizing the 
$n$-dimensional $\mathcal{M}_{\mathbb{H}}$-eigenvalue spectrum. A direct 
consequence of the previous theorems, is the following corollary.

\noindent {\tt T2 - COROLLARY}

{\em Two $\mathcal{M}_{\mathbb{H}}$ quaternionic eigenvectors with complex
 eigenvalues, $\lambda_{\1}$ and $\lambda_{\2}$, with
 $\lambda_{\2} \neq \lambda_{\1} \neq  \lambda_{\2}^{\ast}$, 
are linearly independent on
$\mathbb{H}$.}

Let
\begin{equation}
\label{C1}
 |\psi_{\la}\rangle (\alpha_{\1} + j\beta_{\1}) +
 |\psi_{\lb}\rangle (\alpha_{\2} + j\beta_{\2})
\end{equation}
be a quaternionic linear combination of such eigenvectors.
By taking the complex translation of eq.~(\ref{C1}), we obtain
\begin{equation}
\label{C2}
\alpha_{\1}|\phi_{\la}\rangle  +\, \beta_{\1}|\phi_{\lac}\rangle  +\,
\alpha_{\2}|\phi_{\lb}\rangle  +\, \beta_{\2}|\phi_{\lbc}\rangle ~.
\end{equation} 
The set of $\widetilde{M}$-eigenvectors
\[
\biggl\{ |\phi_{\la}\rangle ~,~|\phi_{\lac}\rangle ~,~|\phi_{\lb}\rangle ~,~|\phi_{\lbc}\rangle 
\biggr\}
\]
is linear independent on $\mathbb{C}$. In fact,  theorem T1  
ensures linear independence between eigenvectors associated to
 conjugate pairs of eigenvalues, and the condition 
$\lambda_{\2} \neq \lambda_{\1} \neq  \lambda_{\2}^{\ast}$ 
completes the proof by assuring the linear independence between
\[
\biggl\{ |\phi_{\la}\rangle ~,~|\phi_{\lac}\rangle 
\biggr\}~~~~~
\mbox{and}~~~~~
\biggl\{ |\phi_{\lb}\rangle ~,~|\phi_{\lbc}\rangle 
\biggr\}~.
\]
Thus the linear combination in eq.~(\ref{C2}), complex 
counterpart of eq.~(\ref{C1}), is null 
if and only if $\alpha_{\1 , \2} = \beta_{\1 , \2} = 0$, 
and consequently the quaternionic linear eigenvectors
$|\psi_{\la}\rangle $ and $|\psi_{\lb}\rangle $ are linear independent
 on $\mathbb{H}$ $\blacksquare$.

%%%%%%%%%%%%%%%%%%%%%%%%%%%%%%%%%%%%%%%%%%%%%%%%%%%%%%%%%%%%%%%%%%%%%%%%%%%
%%%%%%%%%%%%%%%%%%%%%%%%%%%%%%%%%%%%%%%%%%%%%%%%%%%%%%%%%%%%%%%%%%%%%%%%%%%

\subsection*{$\bullet$ A brief discussion about the spectrum choice} 

What happens to the eigenvalue spectrum when we have two simultaneous 
diagonalizable quaternionic linear operators? We show that for
complex operators the choice of a common quaternionic eigenvector set
reproduce in qQM the standard results of complex Quantum Mechanics [cQM].
Let  
\begin{equation}
\label{a12}
\mathcal{A}_{\1} = 
\begin{pmatrix} 
i & 0 \\
0 & i
\end{pmatrix} \, E ~~~~~\mbox{and}~~~~~                   
\mathcal{A}_{\2} = \frac{\hbar}{2} \,  
\begin{pmatrix} 
i & 0 \\
0 & $-$ i
\end{pmatrix}
\end{equation}
be anti-hermitian complex operators associated respectively to energy and 
spin.
In cQM, the corresponding eigenvalue spectrum is  
\begin{equation}
\label{e12}
\biggl\{ iE~,~iE \biggr\}_{\mathcal{A}_{\1}}~~~~~\mbox{and}~~~~~
\biggl\{ i  \frac{\hbar}{2} ~, \, - i \frac{\hbar}{2} 
\biggr\}_{\mathcal{A}_{\2}}~,
\end{equation}
and physically we can describe a particle with positive energy $E$ and spin 
$\frac{1}{2}$. What happens in qQM with quaternionic geometry? The complex 
operators in eq.~(\ref{a12}) also represent two-dimensional 
quaternionic linear operators and so we can translate them in the complex 
world and then extract the eigenvalue spectrum. By following the method 
given in this section, we find the following eigenvalues 
\[
\biggl\{ i E~,~-i E~,~ i E~,~-i E \biggr\}_{\mathcal{A}_{\1}}~~~~~
\mbox{and}~~~~~
\biggl\{ i \frac{\hbar}{2}~, \, - i \frac{\hbar}{2}~,~ 
         i \frac{\hbar}{2}~, \, - i \frac{\hbar}{2}
\biggr\}_{\mathcal{A}_{\2}}~,
\]
and adopting the positive imaginary part convention we extract
\[
\biggl\{ i E~,~i E~ \biggr\}_{\mathcal{A}_{\1}}~~~~~\mbox{and}~~~~~
\biggl\{ i \frac{\hbar}{2}~,~ i \frac{\hbar}{2} 
\biggr\}_{\mathcal{A}_{\2}}~.
\]
It seems that we lose the physical meaning of $\frac{1}{2}$-spin  
particle with positive energy. How can we recover the different sign in
the spin eigenvalues? The solution to this apparent puzzle is represented
by the choice of a {\em common} quaternionic eigenvector set. In fact,
we observe that the previous eigenvalue spectra are related to the following
eigenvector sets
\[
\left\{ 
\begin{pmatrix} 1 \\ 0 \end{pmatrix}~,~
\begin{pmatrix} 0 \\ 1 \end{pmatrix} 
\right\}_{\mathcal{A}_{\1}}
~~~~~\mbox{and}~~~~~
\left\{ 
\begin{pmatrix} 1 \\ 0 \end{pmatrix}~,~
\begin{pmatrix} 0 \\ j \end{pmatrix} 
\right\}_{\mathcal{A}_{\2}}~.
\]
By fixing a common set of eigenvectors 
\begin{equation}
\left\{ 
\begin{pmatrix} 1 \\ 0 \end{pmatrix}~,~
\begin{pmatrix} 0 \\ 1 \end{pmatrix} 
\right\}_{\mathcal{A}_{\1 , \2}}
~,
\end{equation}
we recover the standard results of eq.~(\ref{e12}). Obviously,
\begin{eqnarray*}
\left\{ 
\begin{pmatrix} 1 \\ 0 \end{pmatrix}~,~
\begin{pmatrix} 0 \\ 1 \end{pmatrix} 
\right\}_{\mathcal{A}_{\1 , \2}} & ~~\rightarrow~~ &
\biggl\{ + i E~, \, + i E~ \biggr\}_{\mathcal{A}_{\1}}~,~~~
\biggl\{ + i \frac{\hbar}{2}~, \,  -i \frac{\hbar}{2} 
\biggr\}_{\mathcal{A}_{\2}}~,\\
\left\{ 
\begin{pmatrix} 1 \\ 0 \end{pmatrix}~,~
\begin{pmatrix} 0 \\ j \end{pmatrix} 
\right\}_{\mathcal{A}_{\1 , \2}} & ~~\rightarrow~~ &
\biggl\{ + i E~, \, - i E~ \biggr\}_{\mathcal{A}_{\1}}~,~~~
\biggl\{ + i \frac{\hbar}{2} ~, \,  +i \frac{\hbar}{2}
\biggr\}_{\mathcal{A}_{\2}}~,\\
\left\{ 
\begin{pmatrix} j \\ 0 \end{pmatrix}~,~
\begin{pmatrix} 0 \\ 1 \end{pmatrix} 
\right\}_{\mathcal{A}_{\1 , \2}} & ~~\rightarrow~~ &
\biggl\{ - i E~, \, + i E~ \biggr\}_{\mathcal{A}_{\1}}~,~~~
\biggl\{ - i \frac{\hbar}{2}~, \, -i \frac{\hbar}{2}
\biggr\}_{\mathcal{A}_{\2}}~,\\
\left\{ 
\begin{pmatrix} j \\ 0 \end{pmatrix}~,~
\begin{pmatrix} 0 \\ j \end{pmatrix} 
\right\}_{\mathcal{A}_{\1 , \2}} & ~~\rightarrow~~ &
\biggl\{ - i E~, \,- i E~ \biggr\}_{\mathcal{A}_{\1}}~,~~~
\biggl\{ - i \frac{\hbar}{2}~, \, +i \frac{\hbar}{2} 
\biggr\}_{\mathcal{A}_{\2}}~,
\end{eqnarray*}
represent equivalent choices. Thus, in this particular case, the different 
possibilities in choosing our quaternionic eigenvector set will give the 
following outputs
\begin{center}  
\begin{tabular}{llcll}
Energy : & 
$~~~+E~,~+E$ & ~~~~~and~~~~~ & 
$\frac{1}{2}$-spin : &
$~~~\uparrow~,~\downarrow$ \\
  & 
$~~~+E~,~-E$ & ~~~~~and~~~~~ & 
  &
$~~~\uparrow~,~\uparrow$ \\
  & 
$~~~-E~,~+E$ & ~~~~~and~~~~~ & 
  &
$~~~\downarrow~,~\downarrow$ \\
  & 
$~~~-E~,~-E$ & ~~~~~and~~~~~ & 
  &
$~~~\downarrow~,~\uparrow$~ 
\end{tabular}
\end{center}
Thus, we can also describe a $\frac{1}{2}$-spin particle with positive energy
by re-interpretating spin up/down negative energy as
spin down/up positive energy solutions
\[ 
- E~,~\uparrow \, (\downarrow)~~~\rightarrow~~~ 
  E~,~\downarrow\, (\uparrow)~.
\]

%%%%%%%%%%%%%%%%%%%%%%%%%%%%%%%%%%%%%%%%%%%%%%%%%%%%%%%%%%%%%%%%%%%%%%%%%%%
%%%%%%%%%%%%%%%%%%%%%%%%%%%%%%%%%%%%%%%%%%%%%%%%%%%%%%%%%%%%%%%%%%%%%%%%%%%

\subsection*{$\bullet$ From anti-hermitian to hermitian matrix operator} 

Let us  remark an important difference between the structure of 
an anti-hermitian operator in complex and in quaternionic   
Quantum Mechanics.
In cQM, we can always trivially relate an anti-hermitian
operator, $\mathcal{A}$ to an hermitian operator, $\mathcal{H}$, by 
removing a factor $i$
\[ \mathcal{A} = i \, \mathcal{H}~.\]
In qQM, we must take care. For example, 
\begin{equation}
\label{ant}
\mathcal{A} = 
\left( \begin{array}{cc} $-$ i & 3 j \\ 3 j & i \end{array} \right)~,
\end{equation}
is an anti-hermitian operator, nevertheless, $i \mathcal{A}$ does not 
represent an hermitian operator. The reason is simple: Given any independent,
over $\mathbb{H}$, set of normalized eigenvectors $| v_{\l} \rangle$ of 
$\mathcal{A}$ with complex imaginary eigenvalues $\lambda_{\l}$,
\[ \mathcal{A} = \sum_{l} |v_{\l} \rangle  \, | \lambda_{\l} | i \, 
\langle v_{\l} |~,\]
the corresponding hermitian operator $\mathcal{H}$ is soon obtained by
\[ \mathcal{H} = \sum_{l} |v_{\l} \rangle  \, | \lambda_{\l} | \, 
\langle v_{\l} |~,\]
since both the factors are independent of the particular representative
$|v_{\l} \rangle$ chosen. 
Due to the non-commutative nature of $ | v_{\l} \rangle $, 
we cannot extract the
complex imaginary unit $i$. Our approach to quaternionic eigenvalues 
equations contains a practical method to find eigenvectors $ | v_{\l} \rangle $
 and eigenvalues $\lambda_{\l}$ and consequently solves the problem 
to determine, given a quaternionic  anti-hermitian operator, the corresponding 
hermitian  operator. An easy computation shows that
\[  
\left\{ i | \lambda_{\1} | \, , \, i | \lambda_{\2} | \right\} = 
\left\{ 2 i \, , \, 4 i \right\}~~~~~\mbox{and}~~~~~
\left\{ | v_{\1} \rangle  \, , \, | v_{\2} \rangle  \right\}
 = 
\left\{ \, \frac{1}{\sqrt{2}} \, \left( \begin{array}{c} i \\ j \end{array} 
\right) \, , \, \frac{1}{\sqrt{2}} \, 
~\left( \begin{array}{c} k \\ 1 \end{array} \right) \, \right\}~.
\]
So, the hermitian operator corresponding to the anti-hermitian operator
of eq.~(\ref{ant}) is 
\begin{equation}
\mathcal{H} = 
\left( \begin{array}{cc} 3 & k \\ $-$ k & 3 \end{array} \right)~.
\end{equation}

%%%%%%%%%%%%%%%%%%%%%%%%%%%%%%%%%%%%%%%%%%%%%%%%%%%%%%%%%%%%%%%%%%%%%%%%%%%
%%%%%%%%%%%%%%%%%%%%%%%%%%%%%%%%%%%%%%%%%%%%%%%%%%%%%%%%%%%%%%%%%%%%%%%%%%%

\subsection*{$\bullet$ A practical rule for diagonalization} 

We know that $2n$-dimensional complex operators,  are diagonalizable 
if and only if they admit $2n$ linear independent eigenvectors. 
It is easy to demonstrate that the diagonalization  
matrix for $\widetilde{M}$ 
\[
\widetilde{S} \,  \widetilde{M} \, 
\widetilde{S}^{\- \1} = 
\widetilde{M}^{\d}~,
\]
is given by
\begin{equation}
\widetilde{S}  = \mbox{\tt Inverse} \, \left[
\begin{pmatrix}
 x_{\1}^{\lae} & x_{\1}^{\laec} & ... & x_{\1}^{\lne} & x_{\1}^{\lnec} \\ 
 y_{\1}^{\lae} & y_{\1}^{\laec} & ... & y_{\1}^{\lne} & y_{\1}^{\lnec} \\
\vdots & \vdots & \ddots & \vdots & \vdots \\ 
 x_{\n}^{\lae} & x_{\n}^{\laec} & ... & x_{\n}^{\lne} & x_{\n}^{\lnec} \\ 
 y_{\n}^{\lae} & y_{\n}^{\laec} & ... & y_{\n}^{\lne} & y_{\n}^{\lnec} \\ 
\end{pmatrix} 
\right]~.
\end{equation}
Such a matrix is in the same subset of $\widetilde{M}$, i.e.   
$\widetilde{S}  \in \widetilde{M}_{\dn} (\mathbb{C})$. In fact, by 
recalling the relationship between $| \phi_{\ll} \rangle $ and $| \phi_{\lc} \rangle $, 
we can rewrite the previous diagonalization matrix as  
\begin{equation}
\label{p}
\widetilde{S}  = \mbox{\tt Inverse} \, \left[
\begin{pmatrix}
 x_{\1}^{\lae} & $-$y_{\1}^{* \, \lae} & ... & 
                                      x_{\1}^{\lne} & $-$y_{\1}^{* \,\lne} \\ 
 y_{\1}^{\lae} & x_{\1}^{* \lae} & ... & 
                                      y_{\1}^{\lne} & x_{\1}^{* \, \lne} \\
\vdots & \vdots & \ddots & \vdots & \vdots \\ 
 x_{\n}^{\lae} & $-$y_{\n}^{* \,\lae} & ... & 
                                      x_{\n}^{\lne} & $-$y_{\n}^{* \, \lne} \\ 
 y_{\n}^{\lae} & x_{\n}^{* \, \lae} & ... & 
                                      y_{\n}^{\lne} & x_{\n}^{* \, \lne} \\ 
\end{pmatrix}
\right]~.
\end{equation}
The linear independence of the $2n$ complex eigenvectors of $\widetilde{M}$ 
guarantees the existence of  
$\widetilde{S}^{\- \1}$ and 
the isomorphism between the group of $n \times n$ invertible quaternionic 
matrices $\mathsf{GL}(n,\mathbb{H})$ and the complex counterpart group  
$\widetilde{\mathsf{GL}}(2n,\mathbb{C})$ ensures 
$\widetilde{S}^{\- \1}  \in \widetilde{M}_{\dn} (\mathbb{C})$.
So, the quaternionic $n$-dimensional
matrix which diagonalizes $\mathcal{M}_{\mathbb{H}}$
\[
\mathcal{S}_{\mathbb{H} } \, \mathcal{M}_{\mathbb{H}} \,  
\mathcal{S}_{\mathbb{H} }^{\- \1} = 
\mathcal{M}_{\mathbb{H}}^{\d}~,
\]
can be directly obtained by translating eq.~(\ref{p}) in 
\begin{equation}
\mathcal{S}_{\mathbb{H}} = \mbox{\tt Inverse} \, \left[
\begin{pmatrix}
 x_{\1}^{\lae} + jy_{\1}^{\lae} & ... & x_{\1}^{\lne} + jy_{\1}^{\lne} \\ 
\vdots & \ddots & \vdots \\ 
 x_{\n}^{\lae} + jy_{\n}^{\lae} & ... & x_{\n}^{\lne} + jy_{\n}^{\lne} \\ 
\end{pmatrix}
\right]~.
\end{equation}
In translating complex matrices in quaternionic language, we remember that
an appropriate mathematical notation should require the use of the
left/right quaternionic operators $L_{i,j,k}$ and $R_{i}$. In this case,
due to the particular form of our complex matrices,  
\[ 
\widetilde{M} \, , ~\widetilde{S}
\, , ~ \widetilde{S}^{\- \1}  \in \widetilde{M}_{\dn} (\mathbb{C})~,
\]
their quaternionic translation is performed by left operators and so we use
the simplified notation $i$, $j$, $k$ instead of 
$L_{i}$,  $L_{j}$,  $L_{k}$.

This diagonalization quaternionic matrix is strictly
related to the choice of a particular set of quaternionic linear independent
eigenvectors 
\[
\biggl\{ | \psi_{\la} \rangle  ~,~...~,~|\psi_{\ln}\rangle  \biggr\}~.
\]
So, the diagonalized quaternionic matrix reads 
\[ \mathcal{M}_{\mathbb{H}}^{\d} = \mbox{diag}~ 
\left\{ \lambda_{\1} ~,~...~,~\lambda_{\n} \right\}~.
\]
The choice of a different quaternionic eigenvector set
\[
\biggl\{ \left[ |\psi_{\la}\rangle ~ \mbox{or} ~|\psi_{\lac}\rangle  \right]~,~...~,~
\left[ |\psi_{\ln}\rangle ~ \mbox{or} ~|\psi_{\lnc}\rangle  \right] \biggr\}~,
\]  
will give, for not real eigenvalues, 
a different diagonalization matrix and consequently a different
diagonalized quaternionic matrix
\[ \mathcal{M}_{\mathbb{H}}^{\d} = \mbox{diag} \,  
\left\{ [ \lambda_{\1}~ \mbox{or} ~\lambda_{\1}^{*} ]~,~...~,~
           [ \lambda_{\n}~ \mbox{or} ~\lambda_{\n}^{*} ]  \right\}~.
\]
In conclusion,
\[
\mathcal{M}_{\mathbb{H}}~~~\mbox{diagonalizable}~~~\Leftrightarrow~~~
\widetilde{M}~~~\mbox{diagonalizable}~,
\]
and the diagonalization quaternionic matrix can be easily obtained from the
quaternionic eigenvector set.

%%%%%%%%%%%%%%%%%%%%%%%%%%%%%%%%%%%%%%%%%%%%%%%%%%%%%%%%%%%%%%%%%%%%%%%%%%%%
%                   SUBSECTION 3.2
%%%%%%%%%%%%%%%%%%%%%%%%%%%%%%%%%%%%%%%%%%%%%%%%%%%%%%%%%%%%%%%%%%%%%%%%%%%%

\subsection{Complex linear operators and complex geometry}

In this section, we discuss right eigenvalue equation for complex linear
operators. In $n$-dimensional quaternionic vector spaces, $\mathbb{H}^{n}$,
complex linear operator, $O_{\mathbb{C}}$, are represented by
$n \times n$ quaternionic matrices, 
$\mathcal{M}_{\n}(\mathbb{H}^{\L} \otimes \mathbb{C}^{\R})$, 
 with elements
in $\mathbb{H}^{\L} \otimes \mathbb{C}^{\R}$. 
Such quaternionic matrices admit $2n$-dimensional 
complex counterparts which recover the {\em full} set of $2n$-dimensional
complex matrices, $M_{\dn}(\mathbb{C})$. It is immediate to check that
quaternionic matrices 
$\mathcal{M}_{\mathbb{H}} \in \mathcal{M}_{\n}(\mathbb{H}^{\L})$ are 
characterized by $4n^2$ real parameters and so a natural translation gets the
complex matrix $\widetilde{M}_{\dn}(\mathbb{C}) \subset M_{\dn}(\mathbb{C})$,
whereas a generic $2n$-dimensional complex matrix $M \in M_{\dn}(\mathbb{C})$,
characterized by $8n^2$ real parameters needs to double the $4n^2$ real 
parameters of $\mathcal{M}_{\mathbb{H}}$. By allowing right-action for the 
imaginary units $i$ we recover the missing real parameters. So, the
$2n$-dimensional complex eigenvalue equation
\begin{equation}
\label{cee}
M | \phi \rangle  \, = \lambda | \phi \rangle  ~,~~~
M \in M_{\dn}(\mathbb{C})~,~
| \phi \rangle  \in \mathbb{C}^{\, \dn}~,~\lambda \in \mathbb{C}~,
\end{equation}
becomes, in quaternionic formalism,
\begin{equation}
\label{qee}
\mathcal{M}_{\mathbb{C}} | \psi \rangle  \, = | \psi \rangle  \lambda~,~~~
\mathcal{M}_{\mathbb{C}} \in 
\mathcal{M}_{\n}(\mathbb{H^{\L}} \otimes \mathbb{C}^{\R})~,~
| \psi \rangle  \, \in \mathbb{H}^{\, \n}~,~\lambda \in \mathbb{C}~.
\end{equation}
The right position of the complex eigenvalue $\lambda$ is justified by the 
translation rule
\[ 
i {\u}_{\dn} ~~~\leftrightarrow ~~~ R_{i} \u_{\n}~.
\]
By solving the complex eigenvalue problem of eq.~(\ref{cee}), we find $2n$ 
eigenvalues and we have no possibilities to classify or characterize
such a complex eigenvalue spectrum . 
Is it possible  to extract a suitable quaternionic eigenvectors
set? What happens when the complex spectrum is characterized by  $2n$ 
{\em different} complex eigenvalues? 
To give satisfactory answers to these questions we 
must adopt a complex geometry~\cite{HOR,REM}. In this case
\[ | \psi \rangle  ~~~\mbox{and}~~~| \psi \rangle  j \]
represent  orthogonal vectors and so we cannot kill the eigenvectors 
$| \psi  \rangle  j$. So,  for $n$-dimensional
quaternionic matrices $\mathcal{M}_{\mathbb{C}}$ we must consider 
the {\em full} eigenvalue spectrum
\begin{equation}
\biggl\{ \lambda_{\1}~,~...~,~ \lambda_{\dn} \biggr\}~.
\end{equation}
The corresponding quaternionic eigenvector set is then given by
\begin{equation}
\biggl\{ | \psi_{\la} \rangle ~,~...~,~ | \psi_{\ldn} \rangle  \biggr\}~,
\end{equation}
which represents the quaternionic translation of the $M$-eigenvector set
\begin{equation}
\biggl\{ | \phi_{\la} \rangle ~,~...~,~ | \phi_{\ldn} \rangle  \biggr\}~.
\end{equation}

In conclusion within a qQM with complex 
geometry~\cite{DR} we find for quaternionic linear operators,
$M_\mathbb{H}$, and complex linear operators, $M_\mathbb{C}$,
a $2n$-dimensional complex eigenvalue spectrum and consequently
$2n$ quaternionic eigenvectors. Let us now give a practical method to 
diagonalize complex linear operators. Complex $2n$-dimensional matrices, $M$,
are diagonalizable if and only if admit $2n$ linear independent eigenvectors.
The diagonalizable matrix can be written in terms of $M$-eigenvectors as 
follows
\begin{equation}
S  = \mbox{\tt Inverse} \, \left[
\begin{pmatrix}
 x_{\1}^{\lae} & x_{\1}^{\lde} & ... & x_{\1}^{\ldnae} & x_{\1}^{\ldne} \\ 
 y_{\1}^{\lae} & y_{\1}^{\lde} & ... & y_{\1}^{\ldnae} & y_{\1}^{\ldne} \\ 
\vdots & \vdots & \ddots & \vdots & \vdots \\ 
 x_{\n}^{\lae} & x_{\n}^{\lde} & ... & x_{\n}^{\ldnae} & x_{\n}^{\ldne} \\ 
 y_{\n}^{\lae} & y_{\n}^{\lde} & ... & y_{\n}^{\ldnae} & y_{\n}^{\ldne} \\ 
\end{pmatrix}
\right]~.
\end{equation}
This matrix admits a quaternionic counterpart~\cite{DEL6} by complex linear 
operators
\begin{equation}
\mathcal{S}_{\mathbb{C} } =  \mbox{\tt Inverse} \, \left[
\begin{pmatrix}
q_{\1}^{ [ \1 , \2 ] } + p_{\1}^{ [ \1 , \2 ]} R_i & ... & 
q_{\1}^{[ \dn \- \1 , \dn]} +  p_{\1}^{[ \dn \- \1 , \dn ]} R_i\\ 
 \vdots & \ddots & \vdots \\ 
q_{\n}^{[ \1 , \2 ]} +  p_{\n}^{ [ \1 , \2 ]} R_i & ... & 
q_{\n}^{[ \dn \- \1 , \dn ]} + p_{\n}^{[ \dn \- \1 , \dn ]} R_i\\ 
\end{pmatrix}
\right]~,
\end{equation}
where
\[
q_{\l}^{[ \m , \n ]} = 
\frac{ x_{\l}^{\lme} + y_{\l}^{\lne \, *} }{2} + j \,  
\frac{ y_{\l}^{\lme} -  x_{\l}^{\lne \, *} }{2}~,
\]  
and
\[
p_{\l}^{[ \m , \n ]} = 
\frac{ x_{\l}^{\lme} - y_{\l}^{\lne \, *} }{2i} + j \,  
\frac{ y_{\l}^{\lme} +  x_{\l}^{\lne \, *} }{2i}~.
\]  
To simplify the notation we use $i$ $j$, $k$ instead of 
$L_i$, $L_j$, $L_k$. The right operator $R_i$ indicates the right action of
the imaginary unit $i$. The diagonalized quaternionic matrix  
reproduces the quaternionic translation of the complex matrix 
\[
M^{\d} = \mbox{diag} \, \left\{ \lambda_{\1}~,~...~,~\lambda_{\dn} \right\}
\]
into 
\begin{equation}
\mathcal{M}_{\mathbb{C}}^{\d} =
\mbox{diag} \, \left\{ 
\frac{\lambda_{\1} + \lambda_{\2}^*}{2} + 
\frac{\lambda_{\1} - \lambda_{\2}^*}{2i} \, R_i  
 ~,~...~,~
\frac{\lambda_{\dn \- \1} + \lambda_{\dn}^*}{2} + 
\frac{\lambda_{\dn \- \1} - \lambda_{\dn}^*}{2i} \, R_i  
\right\}~.
\end{equation}

%%%%%%%%%%%%%%%%%%%%%%%%%%%%%%%%%%%%%%%%%%%%%%%%%%%%%%%%%%%%%%%%%%%%%%%%%%%%
%%%%%%%%%%%%%%%%%%%%%%%%%%%%%%%%%%%%%%%%%%%%%%%%%%%%%%%%%%%%%%%%%%%%%%%%%%%%
%                       SECTION 4
%%%%%%%%%%%%%%%%%%%%%%%%%%%%%%%%%%%%%%%%%%%%%%%%%%%%%%%%%%%%%%%%%%%%%%%%%%%%
%%%%%%%%%%%%%%%%%%%%%%%%%%%%%%%%%%%%%%%%%%%%%%%%%%%%%%%%%%%%%%%%%%%%%%%%%%%%%

\section{Quaternionic eigenvalue equation}
\label{s4}

By working with quaternions we have different possibilities to write 
eigenvalue equations. In fact, in solving such equations, we could
 consider quaternionic or complex, left or right  eigenvalues.
 In this section, we briefly introduce the problematic inherent to
 quaternionic eigenvalue equations and emphasize the main difficulties
 present in  such an approach.

\subsection{Right quaternionic  eigenvalue equation for complex linear 
operators}

As seen in the previous sections, the right eigenvalue equation
for quaternionic linear operators, $\mathcal{O}_{\mathbb{H}}$, reads
\[
\mathcal{M}_{\mathbb{H}} | \tilde{\psi} \rangle  \, = | \tilde{\psi} \rangle   q~,~~~~~
q \in \mathbb{H}~.
\]
Such an equation 
can be converted into a right complex eigenvalue equation by re-phasing 
the quaternionic eigenvalues, $q$,
\[
\mathcal{M}_{\mathbb{H}} | \tilde{\psi} \rangle   u  = 
| \tilde{\psi} \rangle   u \,  \overline{u} q u =
| \tilde{\psi} \rangle    \lambda~,~~~~~\lambda \in \mathbb{C}~.
\] 
This trick fails for complex linear operators. In fact, by discussing right 
quaternionic eigenvalue equations for  complex linear operators, 
\begin{equation}
\mathcal{M}_{\mathbb{C}} | \psi \rangle  \, = 
| \psi \rangle    q ~,
\end{equation}
due to the presence of the right imaginary unit $i$ in   
$\mathcal{M}_{\mathbb{C}}$,  
we cannot apply quaternionic similarity transformations,
\[
\left( \mathcal{M}_{\mathbb{C}} | \tilde{\psi} \rangle \right)    u 
\neq \mathcal{M}_{\mathbb{C}} \left( | \tilde{\psi} \rangle   u \right)~,~~~~~
 u \in \mathbb{H}~.
\]

Within a qQM with complex geometry~\cite{DEL4,DR,DEL6}, 
a generic anti-hermitian 
operator must satisfy
\begin{equation}
\label{int1}
\langle \phi|\mathcal{A}_{\mathbb{C}} \psi\rangle_{\mathbb{C}}  \, =
- \langle \mathcal{A}_{\mathbb{C}} \phi |\psi\rangle_{\mathbb{C}}~.
\end{equation}
We can immediately find a constraint on our
 $\mathcal{A}_{\mathbb{C}}$-eigenvalues by putting in the previous
 equation $|\phi\rangle \, = |\psi\rangle $,
\begin{equation}
\label{int2}
\langle \psi|\psi q_{\spsi} \rangle_{\mathbb{C}}
= - 
\langle \psi q_{\spsi} | \psi\rangle_{\mathbb{C}}~~  
 \Rightarrow ~~q_{\spsi} = i \alpha_{\spsi} + jw_{\spsi}~,
\end{equation}
\[
\alpha_{\spsi} \in \mathbb{R}~,~~~w_{\spsi} \in \mathbb{C}~.
\]
Thus, complex linear anti-hermitian operators, $\mathcal{A}_{\mathbb{C}}$,
will be characterized by purely imaginary quaternions. An important property 
must be satisfied for complex linear anti-hermitian operators, namely 
eigenvectors $|\phi\rangle $ and $|\psi\rangle $ associated to different 
eigenvalues, $q_{\sphi} \neq q_{\spsi}$, have to be orthogonal in 
$\mathbb{C}$. By combining  Eqs.~(\ref{int1},\ref{int2}), we find 
\[
\langle \phi|\psi q_{\spsi} \rangle_{\mathbb{C}} =
\langle q_{\sphi} \phi |\psi\rangle_{\mathbb{C}} ~.
\]
To guarantee the complex orthogonality of the eigenvectors 
$|\phi\rangle $ and $|\psi\rangle $, namely  
$\langle \phi|\psi \rangle_{\mathbb{C}} = 0$, we must require 
a complex projection for the eigenvalues, $(q)_{\mathbb{C}}$,
\[
q_{\spsi,\sphi} \longrightarrow \lambda_{\spsi,\sphi}~\in \mathbb{C}~.
\]
In conclusion, a consistent discussion of  right eigenvalue equations
within a qQM with complex geometry requires 
complex eigenvalues.

\subsection{Left quaternionic eigenvalue equation}

What happens for left quaternionic eigenvalue equations? In solving
such equations for quaternionic and complex linear operators,
\[
\begin{cases}
\mathcal{M}_{\mathbb{H}} | \tilde{\psi} \rangle  \, = 
\tilde{q} \, | \tilde{\psi} \rangle ~, \nonumber \\
\mathcal{M}_{\mathbb{C}} | \tilde{\psi} \rangle  \, = 
\tilde{q} \, | \tilde{\psi} \rangle  ~,~~~~~\tilde{q} \in \mathbb{H}~,
\end{cases}
\]
we have not a systematic way to approach the problem. In this case,
due to the presence of left quaternionic eigenvalues (translated in complex
formalism by  two-dimensional matrices),  the 
translation trick does not apply and so we must solve directly the problem
in the quaternionic world. 

In discussing left quaternionic eigenvalue equations, we underline
the difficulty hidden in diagonalizing such operators. Let us suppose
that the matrix representations of our operators be digonalized by 
a matrix $\mathcal{S}_{\mathbb{H} / \mathbb{C}}$
\[ 
\mathcal{S}_{\mathbb{H}} \, 
\mathcal{M}_{\mathbb{H}} \,
\mathcal{S}_{\mathbb{H}}^{\- \1} =
\mathcal{M}_{\mathbb{H}}^{\d}
~~~~~\mbox{and}~~~~~
\mathcal{S}_{\mathbb{C}} \, 
\mathcal{M}_{\mathbb{C}} \,
\mathcal{S}_{\mathbb{C}}^{\- \1} =
\mathcal{M}_{\mathbb{C}}^{\d}~.
\]
The eigenvalue equation will be modified in
\[
\begin{cases}
\mathcal{M}_{\mathbb{H}}^{\d} \mathcal{S}_{\mathbb{H}} | \tilde{\psi} \rangle  \, = 
\mathcal{S}_{\mathbb{H}}  
\tilde{q}  
\mathcal{S}_{\mathbb{H}}^{\- \1} \,
\mathcal{S}_{\mathbb{H}}
 \, | \tilde{\psi} \rangle ~, \nonumber \\
\mathcal{M}_{\mathbb{C}}^{\d} \mathcal{S}_{\mathbb{C}} | \tilde{\psi} \rangle  \, = 
\mathcal{S}_{\mathbb{C}}  
\tilde{q}  
\mathcal{S}_{\mathbb{C}}^{\- \1} \,
\mathcal{S}_{\mathbb{C}}  \,
 | \tilde{\psi} \rangle  ~,
\end{cases}
\]
and now, due to the non-commutative nature of $\widetilde{q}$,
\[
\mathcal{S}_{\mathbb{H}/\mathbb{C}}\, \widetilde{q}\,
\mathcal{S}_{\mathbb{H}/\mathbb{C}}^{\- \1} \neq \widetilde{q}~.
\]
So, we can have operators with the same left quaternionic eigenvalues
 spectrum but no similarity transformation relating them.
 This is explicitly show in appendix B, were we discuss examples of 
two-dimensional quaternionic linear operators.
Let us now analyse other difficulties in solving left quaternionic 
eigenvalue equation. Hermitian quaternionic linear operators 
satisfy
\[
\langle \widetilde{\phi}| \mathcal{H}_{\mathbb{H}} \widetilde{\psi}\rangle \, =
\langle \mathcal{H}_{\mathbb{H}} \widetilde{\phi} 
|\widetilde{\psi}\rangle ~.
\]
By putting $|\widetilde{\phi}\rangle  = |\widetilde{\psi}\rangle $ in the
 previous equation we find 
constraints on the quaternionic eigenvalues $\widetilde{q}$
\[
\langle \widetilde{\psi}|\, \widetilde{q}\, \widetilde{\psi}\rangle  =
\langle \widetilde{q} \, \widetilde{\psi} | \widetilde{\psi}\rangle ~.
\]
From this equation we cannot extract the conclusion that $\tilde{q}$ must be
 real, $\widetilde{q} = \widetilde{q}^{\, \sdag}$. In fact
\[
\langle \widetilde{\psi}|( \widetilde{q} - \widetilde{q}^{\, \sdag})
|\widetilde{\psi}\rangle \,
 = 0
\]
could admit quaternionic solutions for $\tilde{q}$ (see the example in 
appendix B).
 So, the first complication is represented by the possibility to 
find hermitian operators with quaternionic eigenvalues. 
Within a qQM, we can overcome this problem by choosing anti-hermitian operators
to represent observable quantities. In fact,
\[
\langle \widetilde{\phi}| \mathcal{A}_{\mathbb{H}} \widetilde{\psi}\rangle \, 
= -
 \langle \mathcal{A}_{\mathbb{H}} \widetilde{\phi} 
|\widetilde{\psi}\rangle ~,
\]
will imply, for $|\widetilde{\phi}\rangle \, = |\widetilde{\psi}\rangle $,
\begin{equation}
\label{con}
\langle \widetilde{\psi}|( \widetilde{q} + \widetilde{q}^{\, \sdag})
|\widetilde{\psi}\rangle \, = 0~.
\end{equation}
In this case, the real quantity, $\widetilde{q} + \widetilde{q}^{\, \sdag}$,
commutes with $|\widetilde{\psi}\rangle $, and so eq.~(\ref{con}) gives the
constraint 
\[
\widetilde{q} = i \alpha + jw~.
\]
We could work with anti-hermitian operators and choose 
$|\widetilde{q}|$ as observable output. Examples 
of left/right eigenvalue equation for  
two-dimensional anti-hermitian operators will be discussed in 
appendix B. In this appendix, we explicitly show an important difference 
between left and right eigenvalue equation for anti-hermitian operators: 
Left and right eigenvalues can have different absolute values and so cannot 
represent the same physical quantity.

%%%%%%%%%%%%%%%%%%%%%%%%%%%%%%%%%%%%%%%%%%%%%%%%%%%%%%%%%%%%%%%%%%%%%%%%%%%%%%
%%%%%%%%%%%%%%%%%%%%%%%%%%%%%%%%%%%%%%%%%%%%%%%%%%%%%%%%%%%%%%%%%%%%%%%%%%%%%%%
%                      SECTION 5
%%%%%%%%%%%%%%%%%%%%%%%%%%%%%%%%%%%%%%%%%%%%%%%%%%%%%%%%%%%%%%%%%%%%%%%%%%%%%
%%%%%%%%%%%%%%%%%%%%%%%%%%%%%%%%%%%%%%%%%%%%%%%%%%%%%%%%%%%%%%%%%%%%%%%%%%%%%%

\section{Conclusions}

The study undertaken in this paper demonstrates the possibility to 
construct a practical method to diagonalize 
quaternionic and complex linear operators on quaternionic vector spaces.
Quaternionic eigenvalue equations have to be right eigenvalue
equations. As shown in our paper, the choice of a right position for
quaternionic eigenvalues is fundamental in 
searching for a diagonalization method. A left position of 
quaternionic eigenvalues gives unwished surprises. For example, we find  
operators  with the same eigenvalues which are not related by similarity 
transformation, hermitian operators with quaternionic eigenvalues,
etc.

Quaternionic linear operators in $n$-dimensional vector spaces take  
infinite spectra of quaternionic eigenvalues. Nevertheless, 
the {\em complex translation trick} ensures that 
such spectra are related by similarity transformations 
and this gives the possibility to choose $n$ representative 
complex eigenvalues to perform calculations. 
The complete set of  quaternionic eigenvalues 
spectra can be generated from the complex eigenvalue spectrum,
\[
\biggl\{ \lambda_{\1} ~ , ~...~,~ \lambda_{\n} \biggr\}~,
\] 
by quaternionic similarity transformations,
\[
\biggl\{ \overline{u}_{\1} \lambda_{\1} u_{\1} ~ , ~...~,~
\overline{u}_{\n} \lambda_{\n} u_{\n} \biggr\}
\]
Such a symmetry is broken when 
we consider a set of diagonalizable operators. In this case 
the freedom in constructing the eigenvalue spectrum for the first
 operator, and consequently the free choice in determining an  
eigenvectors basis, will fix the eigenvalue spectrum for the 
other operators.

The power of the complex translation trick gives the possibility 
to study general properties for quaternionic and complex linear 
operators. Complex linear operators
play an important role within a qQM with complex geometry 
by reproducing the standard complex results in reduced quaternionic vector 
spaces~\cite{DR}. The method of diagonalization becomes very useful in 
the resolution of quaternionic differential equations~\cite{DD}. Consequently
an immediate application is found in solving the 
Schr\"odinger equation  with quaternionic potentials~\cite{DDS}.  

Mathematical topics to be developed are represented  by the discussion 
of the eigenvalue equation for real linear operators, 
$\mathcal{O}_{\mathbb{R}}$, and by a detailed study of the left eigenvalue
equation. Real linear operators are characterized by  
left and right actions of the quaternionic imaginary units $i,j,k$. The 
translation trick now needs to be applied in the real world and  
so,  for a coherent discussion, it will  require the adoption of a 
real geometry.

%%%%%%%%%%%%%%%%%%%%%%%%%%%%%%%%%%%%%%%%%%%%%%%%%%%%%%%%%%%%%%%%%%%%%%%%%%%
%                     ACKNOWLEDEGEMNETS
%%%%%%%%%%%%%%%%%%%%%%%%%%%%%%%%%%%%%%%%%%%%%%%%%%%%%%%%%%%%%%%%%%%%%%%%%%%

\section*{Acknowledgements}
 
The authors wish to express their gratitude to Nir Cohen 
for several helpful comments concerning
quaternionic matrix theory and Jordan form. They are also indebted to
Gisele Ducati   for suggestions   
on possible applications of the diagonalization method 
to quaternionic differential equations and 
for many stimulating conversations. 
G.S. gratefully acknowledges the 
Department of Applied Mathematics, IMECC-Unicamp,  
for the invitation and hospitality. 
This work was partially supported by a fellowship of the Department 
of Physics, Lecce University, (G.S.) and  by a research grant of the 
Fapesp, S\~ao Paulo State, (S.d.L.). Finally, the authors  wish to thank
the referees for drawing attention to interesting references 
and for their remarks which helped to clarify the notation and improve 
the discussion presented in this paper.

%%%%%%%%%%%%%%%%%%%%%%%%%%%%%%%%%%%%%%%%%%%%%%%%%%%%%%%%%%%%%%%%%%%%%%%%%%%

\newpage

%%%%%%%%%%%%%%%%%%%%%%%%%%%%%%%%%%%%%%%%%%%%%%%%%%%%%%%%%%%%%%%%%%%%%%
%                           APPENDIX A
%%%%%%%%%%%%%%%%%%%%%%%%%%%%%%%%%%%%%%%%%%%%%%%%%%%%%%%%%%%%%%%%%%%%%%

\section*{Appendix A\\
Two dimensional right complex eigenvalue equations}

In this appendix we explicitly solve the  right eigenvalues 
equations for quaternionic, 
$\mathcal{O}_{\mathbb{H}}$, and complex 
$\mathcal{O}_{\mathbb{C}}$, linear operators, in two-dimensional quaternionic 
vector spaces.

%%%%%%%%%%%%%%%%%%%%%%%%%%%%%%%%%%%%%%%%%%%%%%%%%%%%%
\subsection*{$\bullet$ Quaternionic linear operators}  
%%%%%%%%%%%%%%%%%%%%%%%%%%%%%%%%%%%%%%%%%%%%%%%%%%%%%%

Let 
\begin{equation}
\label{a1}
\mathcal{M}_{\mathbb{H}}=
\begin{pmatrix}
 i & j  \\ 
 k  & i 
\end{pmatrix}
\end{equation}
be the quaternionic matrix representation associated to a  
quaternionic linear operator in a two-dimensional quaternionic vector space.
Its complex counterpart reads 
\[
\widetilde{M}= 
\begin{pmatrix}
 i & 0  & 0 & $-$1 \\ 
0  & $-$i  & 1  & 0  \\ 
0  & $-$i  & i  & 0  \\ 
$-$i  & 0  & 0  & $-$i   
\end{pmatrix}~.
\]
In order to solve the right eigenvalue problem
\[
 \mathcal{M}_{\mathbb{H}} | \psi \rangle  \, = | \psi \rangle  \lambda~,~~~~~
\lambda \in \mathbb{C}~,
\]
let us  determine the $\widetilde{M}$-eigenvalue spectrum. From 
the constraint 
\[
\mbox{det} \left[ \widetilde{M} - \lambda {\u}_{\4} \right] = 0~,
\]
we find for the  $\widetilde{M}$-eigenvalues the following solutions
\[
\biggl\{ \lambda_{\1} ~ , ~ \lambda_{\1}^{*} ~ , ~
         \lambda_{\2} ~ , ~ \lambda_{\2}^{*} \biggr\}_{\widetilde{M}} = ~
\biggl\{  2^{\frac{1}{4}}e^{i \frac{3}{8} \pi}  ~ , ~ 
 2^{\frac{1}{4}}e^{\- i \frac{3}{8} \pi}  ~ , \, 
 -2^{\frac{1}{4}}e^{\- i \frac{3}{8} \pi}  ~ , \, 
  -2^{\frac{1}{4}}e^{i \frac{3}{8} \pi}  \biggr\}_{\widetilde{M}}~.  
\]
The $\widetilde{M}$-eigenvector set is given by 
\[
\left\{ 
\begin{pmatrix} -1 + i \lambda_{\1} \\ 0 \\ 0 \\ 1
\end{pmatrix}
~,~  
\begin{pmatrix}
 0 \\ -1- i \lambda_{\1}^{*} \\ -1 \\ 0
\end{pmatrix}
~,~ 
\begin{pmatrix} 0 \\ 1 - i \lambda_{\1}^{*} \\ 1 \\ 0
\end{pmatrix}
~,~ 
\begin{pmatrix}
-1 - i \lambda_{\1} \\ 0 \\ 0 \\ 1
\end{pmatrix}
\right\}_{\widetilde{M}}~.
\]
The $\mathcal{M}_{\mathbb{H}}$-eigenvalue spectrum is soon obtained from that
one of $\widetilde{M}$. For example by adopting the positive imaginary part 
convention we find 
\begin{equation}
\biggl\{ \lambda_{\1} ~ , ~  \lambda_{\2} \biggr\}_{\mathcal{M}_{\mathbb{H}}} 
= ~
\biggl\{  2^{\frac{1}{4}}e^{i \frac{3}{8} \pi}  ~ , ~
  -2^{\frac{1}{4}}e^{ \- i \frac{3}{8} \pi}  
\biggr\}_{\mathcal{M}_{\mathbb{H}}}~,  
\end{equation}
and the corresponding quaternionic eigenvector set, defined up to a right 
complex phase, reads
\begin{equation}
\biggl\{ 
 \begin{pmatrix} -1 + i \lambda_{\1} \\ j 
\end{pmatrix}
~,~
\begin{pmatrix}
 j ( 1 - i \lambda_{\1}^{*}) \\ 1 
\end{pmatrix}
\biggr\}_{\mathcal{M}_{\mathbb{H}}}~.
\end{equation}
The quaternionic matrix which diagonalizes $\mathcal{M}_{\mathbb{H}}$
is
\begin{equation}
\label{AA15}
\mathcal{S}_{\mathbb{H}} = \mbox{\tt Inverse} \, \left[
\begin{pmatrix}
 \, -1+ i \lambda_{\1}~  & ~ j(1 - i \lambda_{\1}^{*}) \,  \\ 
 \, j ~ & ~ 1 \, 
\end{pmatrix}
\right]
=
 - \frac{1}{2\, | \lambda_{\1} |^2 } \, 
\begin{pmatrix}
 \, i \lambda_{\1}^{*} ~  & ~ 
j \left[ i \lambda_{\1}  + | \lambda_{\1} |^2 \right] \,  \\ 
 \, k \lambda_{\1}^{*} ~ & ~ 
i \lambda_{\1} - | \lambda_{\1} |^2   \, 
\end{pmatrix}~.
\end{equation}
As remarked in the paper, we have infinite possibilities of 
diagonalization
\[
\biggl\{\overline{u}_{\1} \lambda_{\1} u_{\1} ~ , ~ 
 \overline{u}_{\2} \lambda_{\2} u_{\2}  \biggr\}~.
\]
Equivalent diagonalized matrices can be obtained from
\[
\mathcal{M}_{\mathbb{H}}^{\d} =
\mbox{diag} \biggl\{ \lambda_{\1} ~ , ~ 
  \lambda_{\2}  \biggr\}
\]
by performing a similarity transformation
\[
\mathcal{U}^{\- \1} \mathcal{M}_{\mathbb{H}}^{\d}\, \mathcal{U} =
 \mathcal{U}^{\sdag} \mathcal{M}_{\mathbb{H}}^{\d}\, \mathcal{U}~,
\]
and
\[
\mathcal{U} = \mbox{diag} \, \biggl\{ u_{\1} ~ , ~ u_{\2}  \biggr\}~.
\]
The diagonalization matrix given in eq.~(\ref{AA15}) becomes 
\[
\mathcal{S}_{\mathbb{H}} ~\rightarrow~ \mathcal{U}^{\sdag}
 \mathcal{S}_{\mathbb{H}}~.
\] 

%%%%%%%%%%%%%%%%%%%%%%%%%%%%%%%%%%%%%%%%%%%%%%%%
\subsection*{$\bullet$ Complex linear operators}
%%%%%%%%%%%%%%%%%%%%%%%%%%%%%%%%%%%%%%%%%%%%%%%%

Let 
\begin{equation}
\label{a4}
\mathcal{M}_{\mathbb{C}}=
\begin{pmatrix}
 -iR_{\i}+j & -kR_{\i}+1  \\ 
 -kR_{\i}-1  & iR_{\i}+j 
\end{pmatrix}
\end{equation}
be the quaternionic matrix representation associated to a  
complex linear operator in a two-dimensional quaternionic vector space.
Its complex counterpart is
\[
M = 
\left(
\begin{array}{rrrr}
 1 & $-$1  & 1 & $-$1 \\ 
1  & $-$1  & $-$1  & 1  \\ 
$-$1  & $-$1  & $-$1  & $-$1  \\ 
$-$1  & $-$1  & 1  & 1   
\end{array}
\right)~.
\]
The right complex eigenvalue problem 
\[
\mathcal{M}_{\mathbb{C}} | \psi \rangle  \, = | \psi \rangle  \lambda~,~~~~~
\lambda \in \mathbb{C}~,
\]
can be solved by determining the $M$-eigenvalue spectrum
\begin{equation}
\biggl\{ \lambda_{\1} ~ , ~ \lambda_{\2} ~ , ~
         \lambda_{\3} ~ , ~ \lambda_{\4} 
\biggr\}_{M / \mathcal{M}_{\mathbb{C}} } =~ 
\biggl\{  2 ~ , \,  -2 ~ , ~2i ~ , \, -2i  
\biggr\}_{M / \mathcal{M}_{\mathbb{C}} }~.  
\end{equation}
Such eigenvalues also determine the $\mathcal{M}_{\mathbb{C}}$-eigenvalues
spectrum. The $\mathcal{M}_{\mathbb{C}}$-eigenvector set is obtained by 
translating the complex $M$-eigenvector set
\[
\left\{ 
\begin{pmatrix} 1 \\ 0 \\ 0 \\ $-$1
\end{pmatrix}
~,~
\begin{pmatrix}
 0 \\ 1 \\ 1 \\ 0
\end{pmatrix}
~,~
\begin{pmatrix} 1 \\ $-$i \\ i \\ 1
\end{pmatrix}
~,~
\begin{pmatrix}
$-$i \\ 1 \\ $-$1 \\ $-$i
\end{pmatrix}
\right\}_{M}~,
\]
in quaternionic formalism
\begin{equation}
\biggl\{  
\begin{pmatrix} 1 \\ $-$j 
\end{pmatrix}
~,~
\begin{pmatrix}
 j \\ 1 
\end{pmatrix}
~,~
\begin{pmatrix} 1+k \\ i+j 
\end{pmatrix}
~,~
\begin{pmatrix}
 j - i \\ k- 1 
\end{pmatrix}
\biggr\}_{\mathcal{M}_{\mathbb{C}}}~.
\end{equation}
The quaternionic matrix which diagonalizes $\mathcal{M}_{\mathbb{C}}$
is
\begin{equation}
\mathcal{S}_{\mathbb{C}} = \mbox{\tt Inverse} \, \left[
\begin{pmatrix}
 \, 1~  & ~1+k \,  \\ 
 \, $-$ j ~ & ~ i+j \, 
\end{pmatrix}
\right]
=
\frac{1}{2} \, 
\begin{pmatrix}
 \, 1~  & ~ j \,  \\ 
 \, \frac{1- k}{2}~ & ~ $-$ \frac{i+j}{2} \, 
\end{pmatrix}~,
\end{equation}
and the diagonalized matrix is given by
\begin{equation}
\label{b5} 
\mathcal{M}_{\mathbb{C}}^{\d} = 2 
\begin{pmatrix}
  $-$iR_{\i}  & 0 \\ 
 0  &  i 
\end{pmatrix}.
\end{equation}
This matrix can be directly obtained from the 
$M / \mathcal{M}_{\mathbb{C}}$ eigenvalue spectrum
by translating, in quaternionic formalism, the matrix
\[
M^{\d} = 
\left(
\begin{array}{rrrr}
 2 &  0     & 0  & 0 \\ 
 0  & $-$2  & 0  & 0  \\ 
 0  & 0     & 2 i  & 0  \\ 
 0  & 0  & 0  & $-$ 2 i   
\end{array}
\right)~.
\] 
It is interesting to note that equivalent diagonalized matrices can be
obtained from $\mathcal{M}_{\mathbb{C}}^{\d}$ in eq.~(\ref{b5}) by the 
similarity transformation 
\[ 
\mathcal{U}^{\sdag} \,  
\mathcal{M}_{\mathbb{C}}^{\d} \,  
\mathcal{U}~.
\]
For example by choosing
\[ 
\mathcal{U} 
=
\begin{pmatrix}
 \,$-$j ~  & ~ 0 \,  \\ 
 \,  0  ~  & ~\frac{1 + k}{\sqrt{2}} \,  \\ 
\end{pmatrix}~,
\]
one find
\begin{equation}
\label{a5} 
\mathcal{M}_{\mathbb{C}}^{\d} ~\rightarrow ~
2 
\begin{pmatrix}
  iR_{\i}  & 0 \\ 
 0  &  j 
\end{pmatrix}~.
\end{equation}

\newpage

%%%%%%%%%%%%%%%%%%%%%%%%%%%%%%%%%%%%%%%%%%%%%%%%%%%%%%%%%%%%%%%%%%%%%%
%                           APPENDIX B
%%%%%%%%%%%%%%%%%%%%%%%%%%%%%%%%%%%%%%%%%%%%%%%%%%%%%%%%%%%%%%%%%%%%%%

\section*{Appendix B\\
Two dimensional left quaternionic eigenvalue equations }

Let us now examine left quaternionic eigenvalue equations
for quaternionic linear operators. 

%%%%%%%%%%%%%%%%%%%%%%%%%%%%%%%%%%%%%%%%%%%%%%%%%%%%%
\subsection*{$\bullet$ Hermitian operators}
%%%%%%%%%%%%%%%%%%%%%%%%%%%%%%%%%%%%%%%%%%%%%%%%%%%%%

Let 
\[
\mathcal{H}_{\mathbb{H}}=
\begin{pmatrix}
 0 & k  \\ 
 $-$k  & 0 
\end{pmatrix}
\]
be the quaternionic matrix representation associated to an 
hermitian quaternionic linear operator.
We consider its left quaternionic eigenvalue equation
\begin{equation}
\label{AB1}
\mathcal{H}_{\mathbb{H}} |\widetilde{\psi} \rangle  \, =
\widetilde{q}\,|\widetilde{\psi} \rangle  ~,
\end{equation}
where
\[
| \widetilde{\psi} \rangle  \, = 
\begin{pmatrix} 
\widetilde{\psi}_{\1}\\ 
\widetilde{\psi}_{\2}
\end{pmatrix}
\, \in \mathbb{H}^{\2}~,~~~
\tilde{q} \in \mathbb{H}~.
\]
Eq.~(\ref{AB1}) can be rewritten by the following quaternionic 
system
\[
\begin{cases}
k \, \widetilde{\psi}_{\2} = 
\tilde{q} \, \widetilde{\psi}_{\1}~, \\
-k \, \widetilde{\psi}_{\1} = 
\tilde{q} \, \widetilde{\psi}_{\2}~.
\end{cases}
\]
The solution is 
\[
\biggl\{ \tilde{q}  \biggr\}_{ \mathcal{H}_{\mathbb{H}}} = ~ 
\biggl\{ z + j \beta    \biggr\}_{\mathcal{H}_{\mathbb{H}}}~,
\]
where
\[
z \in \mathbb{C}~,~~~\beta \in \mathbb{R}~,~~~ |z|^{\2} + \beta^{\2} = 1~.
\]
The $\mathcal{H}_{\mathbb{H}}$-eigenvector set is given by
\[
\left\{ 
\begin{pmatrix}
\widetilde{\psi}_{\1}  \\ $-$k(z + j \beta) \widetilde{\psi}_{\1}
\end{pmatrix}
\right\}_{\mathcal{H}_{\mathbb{H}}}~.
\]
It is easy to verify that in this case 
\[
 <\widetilde{\psi}|(\tilde{q} - \tilde{q}^{\sdag})|\widetilde{\psi}\rangle  = 0 
\]
is verified 
for quaternionic eigenvalues $\tilde{q} \neq \tilde{q}^{\sdag}$.

%%%%%%%%%%%%%%%%%%%%%%%%%%%%%%%%%%%%%%%%%%%%%%%%%%%%%
\subsection*{$\bullet$ Anti-hermitian operators}
%%%%%%%%%%%%%%%%%%%%%%%%%%%%%%%%%%%%%%%%%%%%%%%%%%%%%

Let 
\[
\mathcal{A}_{\mathbb{H}}=
\begin{pmatrix}
 j & i  \\ 
 i  & k 
\end{pmatrix}
\]
be the quaternionic matrix representation associated to an 
anti-hermitian quaternionic linear operator. 
Its right complex spectrum is given by
\[
\biggl\{ \lambda_{\1} ~  , ~
         \lambda_{\2}  \biggr\}_{ \mathcal{H}_{\mathbb{H}}} = ~ 
\biggl\{ i \sqrt{2 - \sqrt{2}} ~ ,
 ~i \sqrt{2 + \sqrt{2}}   \biggr\}_{\mathcal{H}_{\mathbb{H}}}~.
\]  
We now consider the left quaternionic eigenvalue equation
\begin{equation}
\label{AB2}
\mathcal{A}_{\mathbb{H}} |\widetilde{\psi} \rangle  \, =
\widetilde{q}\,|\widetilde{\psi} \rangle  ~,
\end{equation}
where
\[
| \widetilde{\psi} \rangle  \, = 
\begin{pmatrix} 
\widetilde{\psi}_{\1}\\ 
\widetilde{\psi}_{\2}
\end{pmatrix}
\, \in \mathbb{H}^{\,\2}~,~~~
\tilde{q} \in \mathbb{H}~.
\]
By solving the following quaternionic system 
\[
\begin{cases}
j \, \widetilde{\psi}_{\1} + i \, \widetilde{\psi}_{\2} = 
\tilde{q} \, \widetilde{\psi}_{\1}~, \\
i \, \widetilde{\psi}_{\1} + k \, \widetilde{\psi}_{\2}  = 
\tilde{q} \, \widetilde{\psi}_{\2}~,
\end{cases}
\]
we find
\[
\biggl\{ \tilde{q}_{\1}~,~\tilde{q}_{\2} 
 \biggr\}_{ \mathcal{A}_{\mathbb{H}}} = ~ 
\biggl\{ \frac{i}{\sqrt{2}} + \frac{j +k}{2}~,~
  \frac{-i}{\sqrt{2}} + \frac{j + k}{2}  
  \biggr\}_{\mathcal{A}_{\mathbb{H}}}~,
\]
and
\[
\left\{ 
\begin{pmatrix}
\widetilde{\psi}_{\1} \\
 (\frac{1}{\sqrt{2}} + \frac{j + k}{2}) \widetilde{\psi}_{\1}
\end{pmatrix}
~,~
\begin{pmatrix}
\widetilde{\psi}_{\1} \\
 (\frac{-1}{\sqrt{2}} + \frac{j + k}{2}) \widetilde{\psi}_{\1}
\end{pmatrix}
\right\}_{\mathcal{H}_{\mathbb{H}}}~. 
\]
We observe that 
\[
\biggl\{ |\overline{u}_{\1} \lambda_{\1} u_{\1}| =  \sqrt{2 - \sqrt{2}} ~  , ~
        |\overline{u}_{\2} \lambda_{\2} u_{\2} = \sqrt{2 + \sqrt{2}}
  \biggr\}
\]  
and
\[
\biggl\{ |\tilde{q}_{\1}| = 1 ~  , ~
        |\tilde{q}_{\2}| = 1
  \biggr\}~.
\]  
Thus, left and right eigenvalues cannot associated to the same physical 
quantity.

%%%%%%%%%%%%%%%%%%%%%%%%%%%%%%%%%%%%%%%%%%%%%%%%%%%%%
\subsection*{$\bullet$ A new possibility}
%%%%%%%%%%%%%%%%%%%%%%%%%%%%%%%%%%%%%%%%%%%%%%%%%%%%%

In order to complete our discussion let us discuss for the  
quaternionic linear operator given in eq.~(\ref{a1}) its left 
quaternionic eigenvalue equation
\begin{equation}
\label{b1}
 \mathcal{M}_{\mathbb{H}} | \widetilde{\psi} \rangle  \, = 
\tilde{q} \,| \widetilde{\psi} \rangle  
~,
\end{equation}
where
\[
| \widetilde{\psi} \rangle  \, = 
\begin{pmatrix} 
\widetilde{\psi}_{\1}\\ 
\widetilde{\psi}_{\2}
\end{pmatrix}
\, \in \mathbb{H}^{\,\2}~,~~~
\tilde{q} \in \mathbb{H}~.
\]
Eq.~(\ref{b1}) can be rewritten by the following  quaternionic system 
\[
\begin{cases}
i \, \widetilde{\psi}_{\1} + j \, \widetilde{\psi}_{\2} = 
\tilde{q} \, \widetilde{\psi}_{\1}~, \\
k \, \widetilde{\psi}_{\1} + i \, \widetilde{\psi}_{\2} = 
\tilde{q} \, \widetilde{\psi}_{\2}~.
\end{cases}
\]
The solution gives for the quaternionic eigenvalue spectrum
\begin{equation}
\biggl\{ \tilde{q}_{\1} ~  , ~
         \tilde{q}_{\2}  \biggr\}_{ \mathcal{M}_{\mathbb{H}}} = ~ 
\biggl\{ i+ \frac{j+k}{\sqrt{2}} ~ ,
 ~ i- \frac{j+k}{\sqrt{2}}    \biggr\}_{\mathcal{M}_{\mathbb{H}}}~,  
\end{equation}
and for the eigenvector set
\begin{equation} 
\left\{ 
\begin{pmatrix}
1 \\ \frac{1-i}{\sqrt{2}}
\end{pmatrix}
~,~
\begin{pmatrix}
1 \\ \frac{i-1}{\sqrt{2}}
\end{pmatrix}
\right\}_{\mathcal{M}_{\mathbb{H}}}~.  
\end{equation}
Let us now consider the following quaternionic linear operator
\begin{equation}
\label{b2} 
\mathcal{N}_{\mathbb{H}} =
\begin{pmatrix}
  i+ \frac{j+k}{\sqrt{2}}  & 0 \\ 
 0  &  i- \frac{j+k}{\sqrt{2}}
\end{pmatrix}~.
\end{equation}
This operator represents a diagonal operator and has the same left 
quaternionic eigenvalue spectrum of 
$\mathcal{M}_{\mathbb{H}}$, notwithstanding
such an operator is not equivalent to 
$\mathcal{M}_{\mathbb{H}}^{\d}$. In fact,
the  $\mathcal{N}_{\mathbb{H}}$-complex counterpart  
is characterized by the following 
eigenvalue spectrum
\[
\biggl\{ i \sqrt{2}~, \, - i \sqrt{2}~, ~
 i \sqrt{2}~, \, - i \sqrt{2}   \biggr\}_{N}~,  
\]
different from the eigenvalue spectrum of the $M$-complex counterpart of
$\mathcal{M}_{\mathbb{H}}$. Thus, there is no similarity transformation
which relates these two operators in the complex world and consequently
by translation there is no a quaternionic matrix which relates 
$\mathcal{N}_{\mathbb{H}}$ to
 $\mathcal{M}_{\mathbb{H}}$.
So, in the quaternionic world, we can have quaternionic linear operators
which have a same left quaternionic eigenvalue spectrum but not related
by a similarity transformation.

%%%%%%%%%%%%%%%%%%%%%%%%%%%%%%%%%%%%%%%%%%%%%%%%%%%%%%%%%%%%%%%%%%%%%%%%%%%%%%%
%                                REFERENCES
%%%%%%%%%%%%%%%%%%%%%%%%%%%%%%%%%%%%%%%%%%%%%%%%%%%%%%%%%%%%%%%%%%%%%%%%%%%%%%%

\end{document}